\documentclass{aa}
\usepackage{babel}
\usepackage{graphicx}
\usepackage{newtxtext,newtxmath}
\usepackage{lipsum}
\usepackage{lscape}
\usepackage{placeins}
\usepackage{color}
\usepackage{textcomp}
\usepackage{amsmath}
\usepackage{longtable}

\newcommand{\prodigy}{\textsc{PRODIGY}}
\newcommand{\modot}{\,\mathrm{M}_\odot}
\usepackage{xcolor}
\usepackage[
    colorlinks=true,
    citecolor=blue,
    linkcolor=blue,
    urlcolor=blue
]{hyperref}

\begin{document}

\title{\prodigy: Identifying the Progenitors of Present-day Galaxy Clusters}

\author{
    Jonathan S. G\'omez\inst{\ref{Madrid}, \ref{CIAFF}}
    \and Daniel Ceverino\inst{\ref{Madrid}, \ref{CIAFF}}
    \and Tomas Hough\inst{\ref{Madrid}, \ref{CIAFF}, \ref{IAP}}
    \and Gustavo Yepes\inst{\ref{Madrid}, \ref{CIAFF}}
}

\institute{
    Departamento de F\'isica Te\'orica, M\'odulo 8, Facultad de Ciencias, Universidad Aut\'onoma de Madrid, 28049 Madrid, Spain. \label{Madrid}\\
    \email{j.s.gomez.u@gmail.com}
    \and
    Centro de Investigaci\'on Avanzada en F\'isica Fundamental (CIAFF), Facultad de Ciencias, Universidad Aut\'onoma de Madrid, 28049 Madrid, Spain. \label{CIAFF}
    \and
    Instituto de Astrof\'isica de La Plata (CCT La Plata, CONICET, UNLP), Paseo del Bosque s/n, La Plata, Argentina. \label{IAP}
}

\abstract{
    Galaxy protoclusters are the extended, anisotropic progenitors of present-day galaxy clusters, making their identification in projected galaxy distributions sensitive to tracer selection and spatial scale. We introduce \prodigy, a two-step algorithm with publicly available source code, designed to recover both extended protocluster regions and compact cores. We calibrate it on the hydrodynamical galaxy catalogues of \textsc{The Three Hundred} Project at $z=1$--4 for stellar-mass thresholds $M_\star>10^8$, $10^9$, and $10^{10}\modot$. We examine each simulated region in three orthogonal projections to assess viewing-direction effects. Tracing galaxies to their $z=0$ descendants defines separate reference populations for the extended environment and the core. Step~1 combines projected nearest-neighbour densities, an overdensity threshold, and friends-of-friends-like connectivity, while Step~2 uses an adaptive linking length derived from internal galaxy separations. We optimize both steps for purity and completeness under a projected-area constraint. Across the three tracer selections, Step~1 completeness (purity) decreases from approximately 90\% (50\%) at $z=1$ to 70\% (25\%) at $z=4$. Step~2 core completeness (purity) spans approximately 85--90\% (65--80\%) at $z=1$ and 85--100\% (60--100\%) at $z=4$. Step~1 is only weakly sensitive to the number of nearest neighbours over $5\leq n_{\rm n}\leq10$. Performance also shows limited dependence on viewing direction compared with region-to-region scatter. The selected areas closely follow those of their corresponding reference populations, allowing extended environments and compact cores to be characterized on distinct spatial scales. These scales motivate complementary follow-up strategies: mapping the extended progenitor environment and prioritising compact cores for spectroscopy. Comparisons with published observational footprint proxies illustrate this distinction. \prodigy\ provides a simulation-calibrated framework for interpreting projected protocluster candidates; direct applications to surveys will require survey-specific calibration. %
}


\authorrunning{Jonathan S. G\'omez et al.}
\date{Accepted on xx xx, xxxx}%
\maketitle
\nolinenumbers

\section{Introduction}
\label{Sec:Introduction}
Within the hierarchical paradigm of structure formation, present-day galaxy clusters are the final collapsed descendants of the rarest and most massive peaks of the early density field \citep{kravtsov2012}. Their high-redshift progenitors, commonly referred to as galaxy protoclusters, can therefore be defined as the ensemble of galaxies, dark matter haloes, and surrounding large-scale structure that could eventually assemble into a massive cluster-sized halo by $z=0$ \citep{chiang2013,overzier2016}.
However, several studies have addressed that massive halos or galaxies do not necessary stand out at the present time \citep{Remus2023,Onions2025, Baxter2026}. Recently, \citep{Witten2026} make use of TNG-Cluster and TNG300 simulations to show that $\sim64$ per cent of reported protoclusters will likely become group-scale objects ($M_{\rm 200c}<10^{14}\modot$) by $z=0$, partially alleviating the tension between observed protocluster and cluster number densities. %
Unlike virialized clusters, protoclusters are not yet dynamically relaxed systems; instead, they are extended, complex, and often fragmented structures that can span several to tens of comoving megaparsecs and are traced as large-scale overdensities in matter, haloes, or galaxies \citep{chiang2013,muldrew2018}. This evolutionary stage is particularly important because it coincides with the rapid assembly of cluster mass and with the epoch during which a large fraction of the stellar content of present-day cluster galaxies is built up \citep{chiang2015,muldrew2018}. Protocluster environments are therefore expected to imprint early differences between cluster and field galaxies through accelerated mass growth, enhanced or stochastic star formation, preprocessing, and the onset of quenching before galaxies are incorporated into the final virialized cluster \citep{muldrew2018}. As a result, protoclusters provide direct laboratories for studying when and where the galaxies that populate present-day clusters formed, how environmental dependence emerged, and how the extended high-redshift cosmic web is transformed into the compact cluster population observed at low redshift.

Observational searches identify overdensities of selected tracers rather than the complete population that will collapse into a cluster at $z=0$. These include Ly$\alpha$ emitters, Lyman-break galaxies, H$\alpha$ emitters, massive or quiescent galaxies, submillimetre galaxies (SMGs), radio galaxies, quasars and AGN, and Ly$\alpha$ forest tomography \citep{castignani2014,chiang2014,chiang2015,toshikawa2018,calvi2023,brinch2023,zhang2026}. UV and emission-line selections favour active star formation, SMGs trace dusty starbursts, stellar-mass cuts trace massive-system assembly, and beacon searches target rare-object environments \citep{castignani2014,calvi2023,vicentin2025a,zhang2026}. Targeted and wide-area searches have found candidates over broad ranges of redshift and depth \citep{chiang2014,toshikawa2018,li2022,gully2024,vicentin2025b}. Euclid, LSST, HSC, PFS, and JWST programmes \citep{laureijs2011,ivezic2019,aihara2018,takada2014,gardner2006} are expected to greatly increase candidate samples, making identification a statistical problem \citep{gully2024,bohringer2025,li2025}.

A detected overdensity can represent a dense core \citep{vicentin2025a,vicentin2025b}, a fragment of a larger structure, projected components, or an association without a single massive descendant \citep{muldrew2015}. Its descendant mass, membership, extent, and contamination must therefore be inferred statistically using simulations or empirical calibration \citep{chiang2013,chiang2014,brinch2023,bohringer2025}. Such inferences are challenging because massive structures at high redshift follow diverse evolutionary histories and may ultimately evolve into galaxy groups rather than massive clusters \citep{Cui2020,Onions2025,Witten2026}. Moreover, identifying only the dense core may miss a substantial fraction of the extended progenitor population \citep{muldrew2015}. This motivates characterising both the compact core and the surrounding protocluster environment. The adopted geometry and aperture are therefore consequential: \citet{lovell2018} identify $\sim10$ cMpc as a useful purity--completeness compromise, while showing that the protocluster probability depends sensitively on the selection scale. Simulations connect such observable selections with physically defined progenitors of present-day clusters.

Protoclusters lack the compact distribution, unique centre and radius, and relaxed intracluster medium typical of low-redshift clusters. Their measured significance, size, and membership depend on tracer selection, smoothing scale, aperture, and redshift interval. Projection, incompleteness, survey selection, and redshift uncertainties can dilute true overdensities or connect unrelated structures, particularly in photometric samples with large line-of-sight depths \citep{castignani2014,chiang2014,lovell2018,gully2024,bohringer2025}. The full-physics hydrodynamical simulations of {\sc The Three Hundred} Project \citep{cui2022:gizmo} provide a controlled setting in which galaxies can be traced to their descendants, defining a reference population unavailable in observations. We use its high-resolution galaxy catalogues, first presented in \citet{gomez2025} and calibrated following \citet{cui2022:gizmo}, to connect resolved galaxy distributions with observationally relevant selections.

We introduce \prodigy, PROtocluster Discovery and Identification algorithm calibrated on simulated GalaxY clusters, which combines nearest-neighbour densities, overdensity thresholds, and FoF-like connectivity in two steps. Step~1 recovers the extended progenitor environment; Step~2 extracts a compact, robust core for membership assignment and follow-up. These are different targets: the extended region traces the full progenitor structure, while the core is less affected by outer boundaries, fragmentation, and projection. This distinction is motivated by simulations in which substantial protocluster mass spans tens of comoving Mpc and the main halo contains only a minor fraction of members at $z\sim2$ \citep{muldrew2015}.

We calibrate both steps against populations defined by tracing galaxies to their $z=0$ clusters. Purity, completeness, their arithmetic mean (meanPC), and the candidate-to-truth area ratio measure recovery, contamination, and footprint agreement. This tests not only the association of an overdensity with a protocluster, but how much of its physical progenitor population is recovered. \prodigy\ is therefore a two-level calibrated framework rather than simply a FoF group finder \citep{chiang2013,lovell2018,gully2024}.

Section~\ref{Sec:Dataset} describes the data and reference populations; Sect.~\ref{Sec:AnalysisMethods} defines \prodigy\ and its calibration. Section~\ref{Sec:Results} reports membership and area recovery, calibration dependences, and the comparison with observational footprints. Section~\ref{Sec:Discussion} discusses earlier approaches, observational implications, and limitations, and Sect.~\ref{Sec:SummaryConclusions} gives our conclusions. Appendices~\ref{App:Step1Slices}--\ref{App:Step2Diagnostic} retain detailed calibration diagnostics; Appendices~\ref{App:CatalogueDetails} and \ref{App:ObservationalFootprints} document catalogue details and the observational compilation.

\section{Dataset}
\label{Sec:Dataset}

\subsection{The Three Hundred simulations}
\label{Sec:Dataset:The300}
All results in this work use exclusively the 7K-GIZMO hydrodynamical galaxy catalogues of {\sc The Three Hundred} Project\footnote{\url{https://www.the300-project.org}.} \citep{cui2018:the300}. The project follows massive clusters and their environments through zoom-in re-simulations selected from the dark-matter-only MultiDark Planck 2 simulation \citep[MDPL2;][]{klypin2016}. The parent box has side length $1\,h^{-1}{\rm Gpc}$ and contains $3840^3$ dark matter particles of mass $1.5\times10^9\,h^{-1}\mathrm{M}_\odot$. It adopts a flat $\Lambda$CDM cosmology consistent with \textit{Planck} \citep{planckcollaboration2016}: $\Omega_{\rm m}=0.307$, $\Omega_{\rm b}=0.048$, $\Omega_\Lambda=0.693$, $h=0.678$, $\sigma_8=0.823$, and $n_{\rm s}=0.96$. The 324 most massive haloes at $z=0$, with $M_{200}\gtrsim6.24\times10^{14}\,h^{-1}\mathrm{M}_\odot$, define the zoom-region centres. The \prodigy\ calibration is therefore anchored to progenitors of massive present-day clusters, not low-mass groups or arbitrary overdensities.

The zoom initial conditions were generated by tracing to $z=120$ all particles within a comoving radius of $15\,h^{-1}{\rm Mpc}$ around each selected $z=0$ cluster. These Lagrangian regions were resampled with high-resolution dark matter and gas, surrounded by progressively lower-resolution particles to preserve the large-scale tidal field. The hydrodynamical framework uses {\sc Gizmo-Simba} \citep{cui2022:gizmo}, the SIMBA galaxy-formation model \citep{Dave2019}, and the updated SIMBA-C implementation \citep{Hough2023}. It includes cooling, star formation, stellar feedback, enrichment, black-hole growth, and AGN feedback, with parameters recalibrated for the cluster regime. The calibration uses low-redshift cluster stellar observables: the stellar mass fraction within $R_{500}$, the BCG stellar mass--halo mass relation, and the satellite stellar mass function \citep{yang2018}.

The earlier 3K hydrodynamical realizations provide the development context but are not an analysis sample here. Their subhalo resolution limit of approximately $10^{11}\,h^{-1}\mathrm{M}_\odot$ motivates the higher-resolution 7K catalogues, developed and calibrated using the same methodology \citep{gomez2025}. For 7K, $m_{\rm DM}=1.8\times10^8\,h^{-1}\mathrm{M}_\odot$ and $m_{\rm gas}=3.0\times10^7\,h^{-1}\mathrm{M}_\odot$, with dark matter haloes resolved down to approximately $10^{10}\,h^{-1}\mathrm{M}_\odot$. This extends the available tracer population towards less massive galaxies relevant to deep surveys such as {\sc Euclid}. Improved resolution does not by itself establish a direct observational comparison: tracer selection, completeness, and survey effects require the separate modelling discussed in Sect.~\ref{Sec:Discussion:Limitations}. Catalogue construction and the stellar-mass selections used here are specified below.

\subsection{Galaxy catalogues and selection}
\label{Sec:Dataset:GalaxySelection}
In this work, we use the 7K-GIZMO hydrodynamical galaxy catalogues of {\sc The Three Hundred} Project. These catalogues contain full three-dimensional positions, velocities, host-halo information, and merger-tree links. However, the \prodigy\ detection itself is performed on projected two-dimensional galaxy distributions; the three-dimensional and merger-tree information is used only to define the descendant-based reference populations and to identify the main progenitor halo. The use of the hydrodynamical sample is important because the galaxies are formed self-consistently within the simulation and can therefore be used directly as tracers of the high-redshift protocluster population. All haloes and subhaloes in the 7K-GIZMO runs were identified with the Amiga Halo Finder\footnote{\url{http://popia.ft.uam.es/AHF}} \citep[AHF;][]{knollmann&knebe2009}, which locates density peaks using an adaptive mesh hierarchy and removes gravitationally unbound particles from each candidate structure. For each halo, AHF provides standard spherical-overdensity quantities such as $R_{200{\rm c}}$ and $M_{200{\rm c}}$, defined respectively as the radius enclosing a mean density equal to 200 times the critical density of the Universe and the total mass enclosed within that radius. In the hydrodynamical catalogues, the properties of the stellar, gas, and dark matter components associated with each bound structure are also recorded.

Halo and subhalo descendants are identified using AHF--{\sc MergerTree}, providing the links required to construct the descendant-defined reference populations in Sect.~\ref{Sec:Dataset:GroundTruth}.

For each simulated galaxy, the catalogue includes intrinsic physical properties such as stellar mass, gas mass, star formation rate, metallicity, position, velocity, and host-halo information.

In this paper, however, the protocluster detector is calibrated using stellar-mass-selected galaxy samples rather than colour- or flux-selected catalogues. Unless otherwise stated, our fiducial tracer population is defined by selecting galaxies with stellar masses above $M_\star > 10^{10}\modot$, while additional samples with thresholds of $M_\star > 10^{9}\modot$ and $M_\star > 10^{8}\modot$ are also considered. These thresholds provide resolved and observationally motivated galaxy samples for detecting overdense regions while limiting the impact of numerical incompleteness near the resolution limit of the 7K-GIZMO catalogues. They also allow us to evaluate the efficiency of \prodigy\ protocluster detection as a function of the adopted stellar-mass cut. The dependence of the \prodigy\ performance on the adopted stellar-mass selection is explored explicitly in Sect.~\ref{Sec:Results}.

\subsection{Redshifts and projected views}
\label{Sec:Dataset:RedshiftsProjections}
We apply \prodigy\ to the 7K-GIZMO galaxy catalogues at the snapshots closest to $z=1$, $z=2$, $z=3$, and $z=4$. These redshifts sample the epoch during which present-day cluster progenitors are still extended, non-virialized structures, while their galaxy populations are sufficiently developed to be traced with stellar-mass-selected samples. At each redshift, we analyse the galaxy distribution in projection in order to mimic the two-dimensional nature of observational searches for overdensities. For each simulated region, three independent projected views are constructed by projecting the galaxy positions onto the $xy$, $xz$, and $yz$ planes. The corresponding line-of-sight directions are therefore the $z$-, $y$-, and $x$-axes, respectively. Unless otherwise stated, \prodigy\ is run independently on each projected view, so that each cluster region contributes three projected realizations at a given redshift.

All positions are expressed in comoving coordinates. Distances measured directly from the simulation outputs are converted from $h^{-1}{\rm Mpc}$ to ${\rm Mpc}$ using the adopted value of $h=0.678$, and are reported as comoving Mpc throughout the paper unless explicitly stated otherwise. Projected separations are computed in the corresponding two-dimensional plane, for example
\begin{equation}
    d_{ij}^{xy} =
    \left[
    \left(x_i-x_j\right)^2 +
    \left(y_i-y_j\right)^2
    \right]^{1/2},
\end{equation}
for the $xy$ projection, with analogous definitions for the $xz$ and $yz$ views. Projected areas are measured in comoving ${\rm Mpc}^2$ and are computed consistently in the same projected coordinate system used by the algorithm. When physical distances or areas are required for comparison, comoving distances are divided by $(1+z)$, and comoving areas by $(1+z)^2$.

The three views quantify projection-induced variance in purity, completeness, meanPC, and area ratio for intrinsically anisotropic protoclusters \citep{matsuda2005,muldrew2015,overzier2016}. This controlled test does not include the line-of-sight selection of a real survey.

\subsection{Ground truth: descendants and real protocluster members}
\label{Sec:Dataset:GroundTruth}
A central advantage of simulations is that protocluster membership can be defined from galaxy descendants rather than inferred from observed overdensities. For each of the 324 zoom regions, the reference $z=0$ cluster corresponds to the central massive halo around which the region was selected. At each analysed redshift, galaxies are classified as protocluster members if their descendants belong to this reference cluster. This defines a descendant-based reference population independent of the projected overdensity selected by PRODIGY. We distinguish the extended protocluster set, $\mathcal{T}^{\rm ext}_{\mathcal{R}}(z)$, and the core set, $\mathcal{T}^{\rm core}_{\mathcal{R}}(z)$.

\subsubsection{Extended protocluster ground truth}
\label{Sec:Dataset:GroundTruth:Extended}

Because \prodigy\ is designed as a two-step algorithm, we use two related but distinct ground-truth definitions. The first one corresponds to the extended protocluster region. For a cluster region $\mathcal{R}$ and redshift $z$, we define
\begin{equation}
    \mathcal{T}^{\rm ext}_{\mathcal{R}}(z) =
    \left\{
    g_i(z) \;\middle|\; D[g_i(z)] \in \mathcal{C}_{\mathcal{R}}(z=0)
    \right\},
\end{equation}
where $g_i(z)$ is a galaxy in the selected stellar-mass-limited tracer sample at redshift $z$, $D[g_i(z)]$ denotes its descendant at $z=0$, and $\mathcal{C}_{\mathcal{R}}(z=0)$ is the final cluster in region $\mathcal{R}$. This set contains all tracer galaxies at redshift $z$ that will become members of the descendant cluster, irrespective of whether they are already located in the main progenitor halo or distributed across neighbouring haloes, groups, and filaments. It is therefore the reference truth used to evaluate the first step of \prodigy, which aims to recover the extended protocluster environment.

\subsubsection{Core ground truth}
\label{Sec:Dataset:GroundTruth:Core}

The second definition corresponds to the protocluster core. At each redshift, we identify the main progenitor halo of the reference $z=0$ cluster as the halo located on the main branch of the corresponding merger tree. We denote this halo by $H^{\rm main}_{\mathcal{R}}(z)$, and define the associated main-progenitor system as

\begin{equation}
    \begin{aligned}
        \mathcal{H}^{\rm main}_{\mathcal{R}}(z)
        ={}&\left\{H^{\rm main}_{\mathcal{R}}(z)\right\}\\
        &\cup\left\{h_j(z):
        h_j(z)\subset H^{\rm main}_{\mathcal{R}}(z),\
        d_j^{\rm main}<R_{200c}^{\rm main}(z)\right\}.
    \end{aligned}
\end{equation}

Here $d_j^{\rm main}$ is the three-dimensional distance between subhalo $h_j$ and the centre of the main progenitor halo, and $R^{\rm main}_{200{\rm c}}(z)$ is the corresponding spherical-overdensity radius. The core truth is then defined as the subset of extended protocluster members whose host halo or subhalo belongs to this main-progenitor system:
\begin{equation}
    \mathcal{T}^{\rm core}_{\mathcal{R}}(z) =
    \left\{
    g_i(z) \in \mathcal{T}^{\rm ext}_{\mathcal{R}}(z)
    \;\middle|\;
    H[g_i(z)] \in \mathcal{H}^{\rm main}_{\mathcal{R}}(z)
    \right\},
\end{equation}
where $H[g_i(z)]$ denotes the halo or subhalo hosting galaxy $g_i$ at redshift $z$. This definition selects the real protocluster members that are already part of the central collapsed or collapsing system around the main progenitor, rather than all galaxies distributed across neighbouring haloes, groups, and filaments. It therefore provides the reference population for Step~2, which is designed to identify the densest and most robust region within the extended protocluster candidate.

\subsubsection{Number of true members}
\label{Sec:Dataset:GroundTruth:Ntruth}

For each cluster region, redshift, stellar-mass threshold, and projected view, we define the number of true members as
\begin{equation}
    N^{\rm ext}_{\rm truth}(\mathcal{R},z) =
    \left|\mathcal{T}^{\rm ext}_{\mathcal{R}}(z)\right|,
    \qquad
    N^{\rm core}_{\rm truth}(\mathcal{R},z) =
    \left|\mathcal{T}^{\rm core}_{\mathcal{R}}(z)\right|.
\end{equation}
The membership of these truth sets is defined in three dimensions and therefore does not depend on the projected view. However, their projected distribution, area, and degree of overlap with the regions selected by \prodigy\ do depend on whether the $xy$, $xz$, or $yz$ projection is used. For this reason, the same three-dimensional truth catalogue is projected onto each of the three orthogonal planes before computing the performance metrics. In the following sections, $N_{\rm truth}$ denotes the appropriate reference number of real members for the corresponding step: $N^{\rm ext}_{\rm truth}$ for the extended-region search and $N^{\rm core}_{\rm truth}$ for the core-identification step.

\section{Analysis methods}
\label{Sec:AnalysisMethods}

\subsection{Overview of the \prodigy{} algorithm}
\label{Sec:Methods:NEXUSOverview}
\prodigy\ is applied independently to each cluster region, redshift, projection, and stellar-mass threshold. The input is the projected galaxy catalogue described in Sect.~\ref{Sec:Dataset}, with stellar masses, host-halo information, and descendant-defined reference membership. Step~1 combines local projected densities, an overdensity threshold, and FoF-like connectivity to select an extended candidate around the projected main progenitor. Step~2 uses the internal spacing of Step~1 members to extract a compact core with an adaptive linking length. The two candidate sets are evaluated against $\mathcal{T}^{\rm ext}_{\mathcal{R}}(z)$ and $\mathcal{T}^{\rm core}_{\mathcal{R}}(z)$, respectively. We measure recovered members, interlopers, missed members, and projected areas, and optimise the parameters using the metrics defined in Sect.~\ref{Sec:Methods:Metrics}.

\subsection{Step 1: extended protocluster candidates}
\label{Sec:Methods:Step1}
Step~1 aims to recover the extended truth population $\mathcal{T}^{\rm ext}_{\mathcal{R}}(z)$ while controlling contamination and projected area. For each stellar-mass-selected tracer $g_i$, let $d_{n_{\rm n},i}$ be the projected distance to its $n_{\rm n}$-th nearest neighbour. We define
\begin{equation}
    \begin{aligned}
        \Sigma_i&=\frac{n_{\rm n}}{\pi d_{n_{\rm n},i}^2},\qquad
        \Sigma_0(z)=\rho_0(z)D^{\rm los}_{\mathcal{R}},\\
        \Delta_i&=\frac{\Sigma_i}{\Sigma_0(z)}
        =\frac{n_{\rm n}}{\pi d_{n_{\rm n},i}^2\rho_0(z)D^{\rm los}_{\mathcal{R}}}.
    \end{aligned}
    \label{Eq:Methods:Step1:Overdensity}
\end{equation}
Here $D^{\rm los}_{\mathcal{R}}$ is the difference between the maximum and minimum galaxy positions along the line-of-sight coordinate. The cosmological reference number density $\rho_0(z)$ follows the mean galaxy number-density evolution in SIMBA \citep{Dave2019}, with the normalization of the original implementation. This external field-like reference avoids estimating the background from cluster-biased zoom regions and is consistent with the GIZMO-SIMBA galaxy-formation model. We use the values at $z=0$, $1$, and $2$ and a log-linear extrapolation of the same trend at $z=3$ and $4$.

For each region and projection, the reference overdensity $\Delta_{\rm ref}$ is the mean $\Delta_i$ of galaxies in the projected annulus around the final cluster centre,
\begin{equation}
    0.9R_{\rm vir}<d_i^{\rm cen}<1.1R_{\rm vir},\qquad
    \Delta_{\rm cut}^{(1)}=\alpha\Delta_{\rm ref},
\end{equation}
where $d_i^{\rm cen}$ is the projected distance to the main progenitor centre and $R_{\rm vir}$ is the descendant-cluster virial radius at $z=0$. Lower $\alpha$ gives more extended and complete but more contaminated selections; higher $\alpha$ gives cleaner, more compact candidates that can miss diffuse components.

Starting at the projected main-progenitor position, we select galaxies with $\Delta_i>\Delta_{\rm cut}^{(1)}$ within distance $\ell_1$. Each newly selected galaxy becomes a seed, and overdense neighbours within $\ell_1$ are added until no further galaxies satisfy both criteria:
\begin{equation}
    C^{(1)}_{\mathcal{R}}(z)={\rm FoF}\left[\Delta_i>\Delta_{\rm cut}^{(1)},\ d_{ij}<\ell_1\right].
\end{equation}
Thus $n_{\rm n}$ controls the density-estimation scale, $\alpha$ the overdensity threshold relative to the $z=0$ reference, and $\ell_1$ the spatial connectivity. The parameter grid, footprints, and performance evaluation are described below.

\subsection{Step 2: robust protocluster cores}
\label{Sec:Methods:Step2}
Step~2 identifies a compact core within the extended Step~1 environment and is evaluated against $\mathcal{T}^{\rm core}_{\mathcal{R}}(z)$, rather than the full extended truth population. For each galaxy in $C^{(1)}_{\mathcal{R}}(z)$, we compute its nearest-neighbour separation within that candidate and set
\begin{equation}
    \begin{aligned}
        d_i^{\rm min}&=\min_{j\ne i}d_{ij},\qquad g_i,g_j\in C^{(1)}_{\mathcal{R}}(z),\\
        \ell_2&=P_{p_2}\left(\{d_i^{\rm min}\}_{g_i\in C^{(1)}_{\mathcal{R}}(z)}\right).
    \end{aligned}
    \label{Eq:Methods:Step2:ll2}
\end{equation}
The percentile $p_2$ is the code parameter `ll2p'. Lower percentiles produce more restrictive cores; higher percentiles retain more of the connected dense structure. If Step~1 selects fewer than two galaxies, we set $\ell_2=\ell_1$ as a fallback.

We then repeat the FoF-like growth from the same projected main-progenitor position, retaining the Step~1 overdensity threshold and replacing $\ell_1$ by $\ell_2$:
\begin{equation}
    C^{(2)}_{\mathcal{R}}(z)={\rm FoF}\left[\Delta_i>\Delta_{\rm cut}^{(1)},\ d_{ij}<\ell_2\right].
\end{equation}
The use of the same overdensity cut ensures that Step 2 identifies a compact subset of the same overdense environment, rather than an independent low-density structure.

By construction, Step 2 is expected to trade some completeness for higher purity and smaller projected area. This motivates its use for membership assignment and efficient follow-up. Its purity, completeness, meanPC, and area ratio are computed relative to the core truth set, as defined in Sect.~\ref{Sec:Methods:Metrics}.

\subsection{Parameter grid and optimisation strategy}
\label{Sec:Methods:ParameterGrid}
We explore the grid in Table~\ref{Table:Methods:ParameterGrid} independently for each stellar-mass threshold, redshift, and projected view, and evaluate every configuration on the cluster regions. Step~1 uses the extended truth set and Step~2 the core truth set. For each metric $Q$, we first calculate the individual-region and projection values, then average at fixed redshift and mass threshold:
\begin{equation}
    \langle Q\rangle=\frac{1}{N_{\rm reg}N_{\rm proj}}
    \sum_{\mathcal{R}=1}^{N_{\rm reg}}\sum_{p=1}^{N_{\rm proj}}Q_{\mathcal{R},p},
\end{equation}
where $N_{\rm reg}$ is the number of analysed regions and $N_{\rm proj}=3$. This applies to purity, completeness, meanPC, and area ratio.

\begin{table}
    \centering
    \caption{Parameter grid explored for the calibration of \prodigy. The parameter $n_{\rm n}$ controls the nearest-neighbour density estimate, $\alpha$ sets the overdensity threshold relative to the $z=0$ reference overdensity, $\ell_1$ is the Step 1 linking length, and $p_2$ is the percentile used to define the adaptive Step 2 linking length.}
    \label{Table:Methods:ParameterGrid}
    \begin{tabular}{lll}
        \hline
        Quantity               & Label       & Values                                         \\
        \hline
        Snapshot redshift      & $z$         & $1,2,3,4$                                      \\
        Projection             & --          & $xy,xz,yz$                                     \\
        Stellar-mass threshold & $M_\star$   & $>10^{8}, >10^{9}, >10^{10}\modot$ \\
        Nearest neighbours     & $n_{\rm n}$ & $5,6,7,8,9,10$                                 \\
        Overdensity factor     & $\alpha$    & $0.1$--$3.0$                                   \\ %
        Step 1 linking length  & $\ell_1$    & $0.1$--$8.0\,h^{-1}{\rm Mpc}$                  \\ %
        Step 2 percentile      & $p_2$       & $1$--$100$                                     \\
        \hline
    \end{tabular}
\end{table}

For Step~2, we fix the best allowed Step~1 parameters and optimise only $p_2$ against the core truth. Unless otherwise stated, the fiducial configuration maximises the ensemble meanPC within the area-based allowed region defined in Sect.~\ref{Sec:Methods:Metrics}. MeanPC gives equal weight to purity and completeness. We also record their separate maxima: purity favours efficient spectroscopic follow-up, while completeness favours recovery of the full environment.

\subsection{Visualisation and area measurements}
\label{Sec:Methods:Areas}
We measure the candidate and truth footprints with two-dimensional convex hulls, without imposing circular symmetry or a fixed aperture. For projected coordinates $\mathbf{x}^{\rm proj}_i=(x^{\rm proj}_i,y^{\rm proj}_i)$ and step $s$, we define

\begin{equation}
    A^{(s)}_{\rm candidates}(\mathcal{R},z)=
    A_{\rm hull}\left[\{\mathbf{x}^{\rm proj}_i\mid g_i\in C^{(s)}_{\mathcal{R}}(z)\}\right],
    \label{Eq:Methods:Areas:CandidateArea}
\end{equation}
and the corresponding truth areas by
\begin{equation}
    A^{(1)}_{\rm truth}(\mathcal{R},z)=
    A_{\rm hull}\left[\{\mathbf{x}^{\rm proj}_i\mid g_i\in\mathcal{T}^{\rm ext}_{\mathcal{R}}(z)\}\right],
    \label{Eq:Methods:Areas:TruthAreaStep1}
\end{equation}
\begin{equation}
    A^{(2)}_{\rm truth}(\mathcal{R},z)=
    A_{\rm hull}\left[\{\mathbf{x}^{\rm proj}_i\mid g_i\in\mathcal{T}^{\rm core}_{\mathcal{R}}(z)\}\right].
    \label{Eq:Methods:Areas:TruthAreaStep2}
\end{equation}
Here $A_{\rm hull}$ denotes convex-hull area; fewer than three unique projected points give zero area. Each footprint is measured in the same projected plane as its membership metrics.

In the implementation, coordinates are in comoving $h^{-1}{\rm Mpc}$ and raw areas in $(h^{-1}{\rm Mpc})^2$. The candidate-to-truth area ratio is unit independent; absolute areas are quoted with explicit units. A convex hull captures the region spanned by irregular member distributions but can include empty space between disconnected components or filaments. We therefore combine area with membership metrics, rather than use it alone, to identify detections that recover many true members only by selecting excessively large regions.

\subsection{Performance metrics}
\label{Sec:Methods:Metrics}
For each region, redshift, projection, and mass threshold, we compare candidate $C^{(s)}_{\mathcal{R}}(z)$ with truth $\mathcal{T}^{(s)}_{\mathcal{R}}(z)$, using the extended set for $s=1$ and the core set for $s=2$. The member counts are
\begin{equation}
    \begin{aligned}
        N^{(s)}_{\rm candidates}&=|C^{(s)}_{\mathcal{R}}(z)|,\qquad
        N^{(s)}_{\rm truth}=|\mathcal{T}^{(s)}_{\mathcal{R}}(z)|,\\
        N^{(s)}_{\rm match}&=|C^{(s)}_{\mathcal{R}}(z)\cap\mathcal{T}^{(s)}_{\mathcal{R}}(z)|.
    \end{aligned}
\end{equation}
Interlopers number $N^{(s)}_{\rm candidates}-N^{(s)}_{\rm match}$, and missed true members $N^{(s)}_{\rm truth}-N^{(s)}_{\rm match}$. Purity, completeness, and their equally weighted mean are
\begin{equation}
    \begin{aligned}
        P^{(s)}&=\frac{N^{(s)}_{\rm match}}{N^{(s)}_{\rm candidates}},\qquad
        C^{(s)}=\frac{N^{(s)}_{\rm match}}{N^{(s)}_{\rm truth}},\\
        {\rm meanPC}^{(s)}&=\frac{P^{(s)}+C^{(s)}}{2}.
    \end{aligned}
\end{equation}
Purity measures contamination control, while completeness measures recovery. Membership alone can favour inefficient detections that cover excessively large regions. We therefore also use
\begin{equation}
    A^{(s)}_{\rm ratio}=\frac{A^{(s)}_{\rm candidates}}{A^{(s)}_{\rm truth}}.
\end{equation}
Ratios below (above) unity indicate smaller (larger) candidate footprints than the corresponding truth. Metrics are averaged over regions and projections as in Sect.~\ref{Sec:Methods:ParameterGrid}; a configuration is \emph{allowed} if $\langle A^{(s)}_{\rm ratio}\rangle\leq1$ and \emph{forbidden} otherwise. Fiducial configurations maximise $\langle{\rm meanPC}\rangle$ within this allowed region, rather than obtain high completeness by overextending the footprint.

If no candidate members are selected, we set $P^{(s)}=C^{(s)}=0$. Cases with fewer than three galaxies in the candidate or truth set have zero convex-hull area and remain in the membership metrics. Cases with zero truth area are excluded from the area-based allowed/forbidden classification.

\section{Results}
\label{Sec:Results}
We present the performance of \prodigy\ in a progressive way. We begin with an illustrative example that highlights the two-step detection strategy and visually compares the extended protocluster truth, the core truth, and the corresponding \prodigy\ detections. We then quantify the global performance of the algorithm across the full set of cluster regions, and finally examine how the results depend on redshift, projected view, stellar-mass threshold, and the parameters controlling the overdensity and connectivity criteria.

\subsection{An illustrative example of \prodigy{} detections}
\label{Sec:Results:Example}
\begin{figure*}
    \centering
    \hspace{-0.4cm} \includegraphics[width=0.37\textwidth]{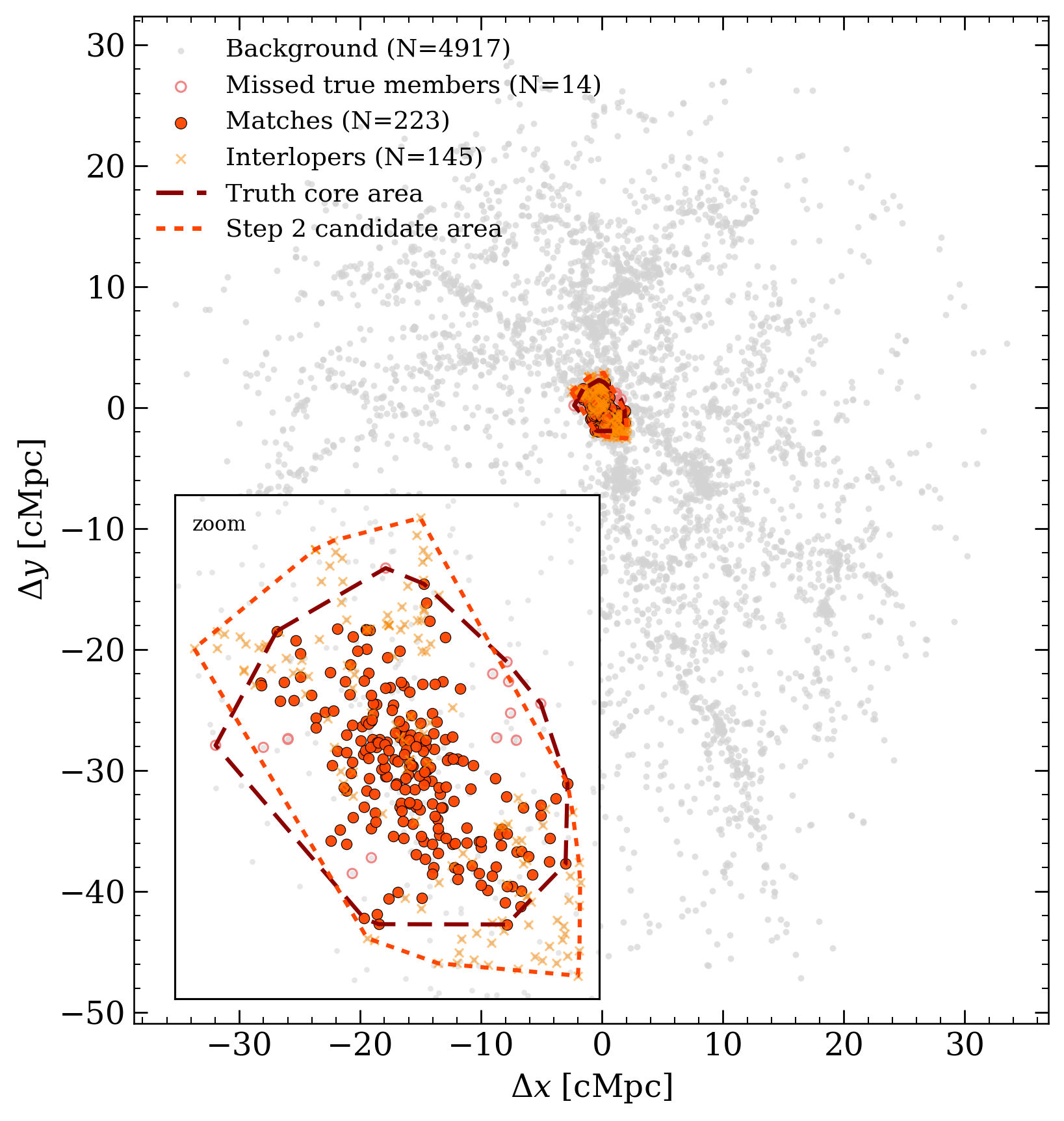}
    \hspace{-12.9cm}\includegraphics[width=0.37\textwidth]{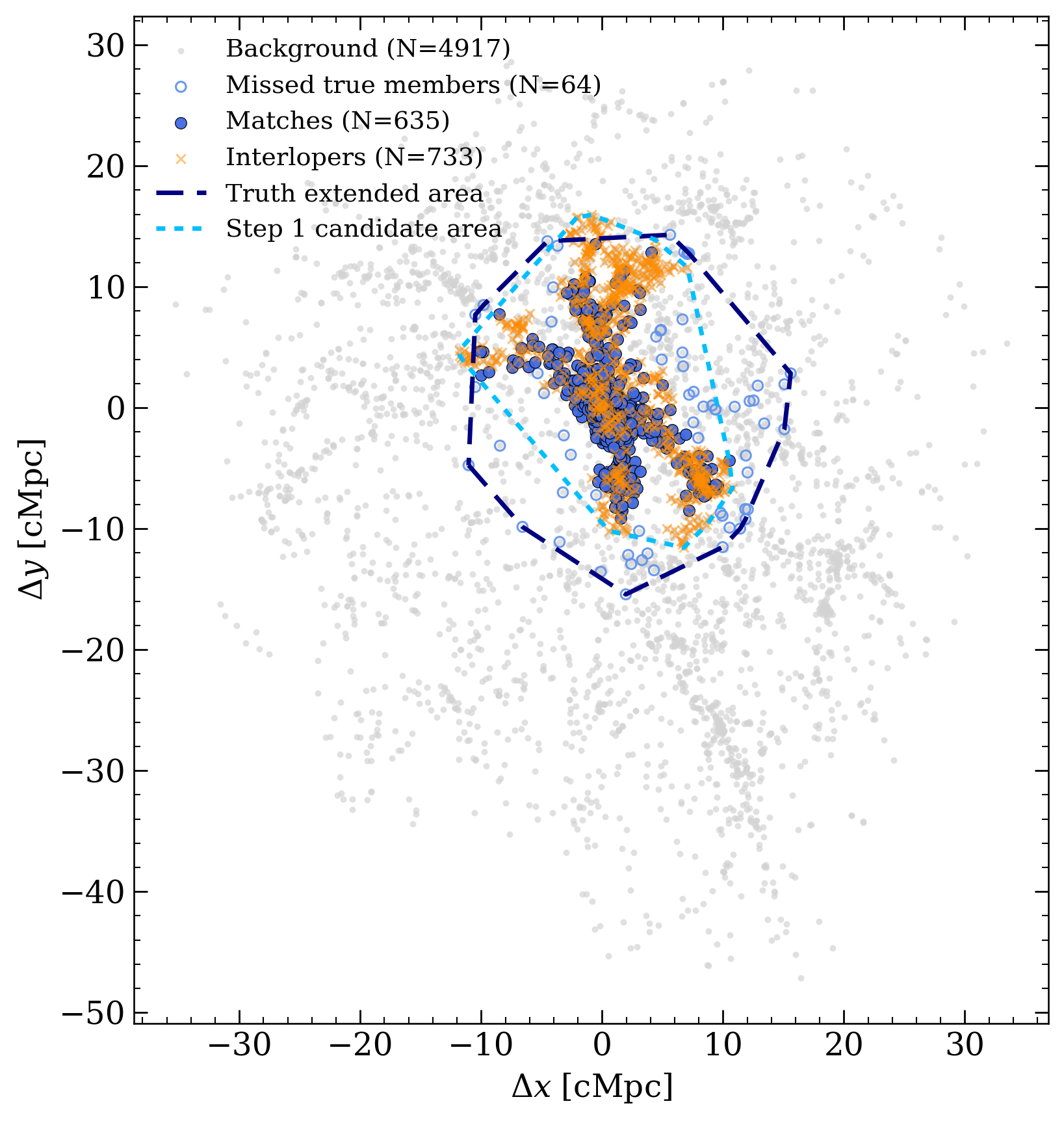}
    \hspace{-12.9cm}\includegraphics[width=0.37\textwidth]{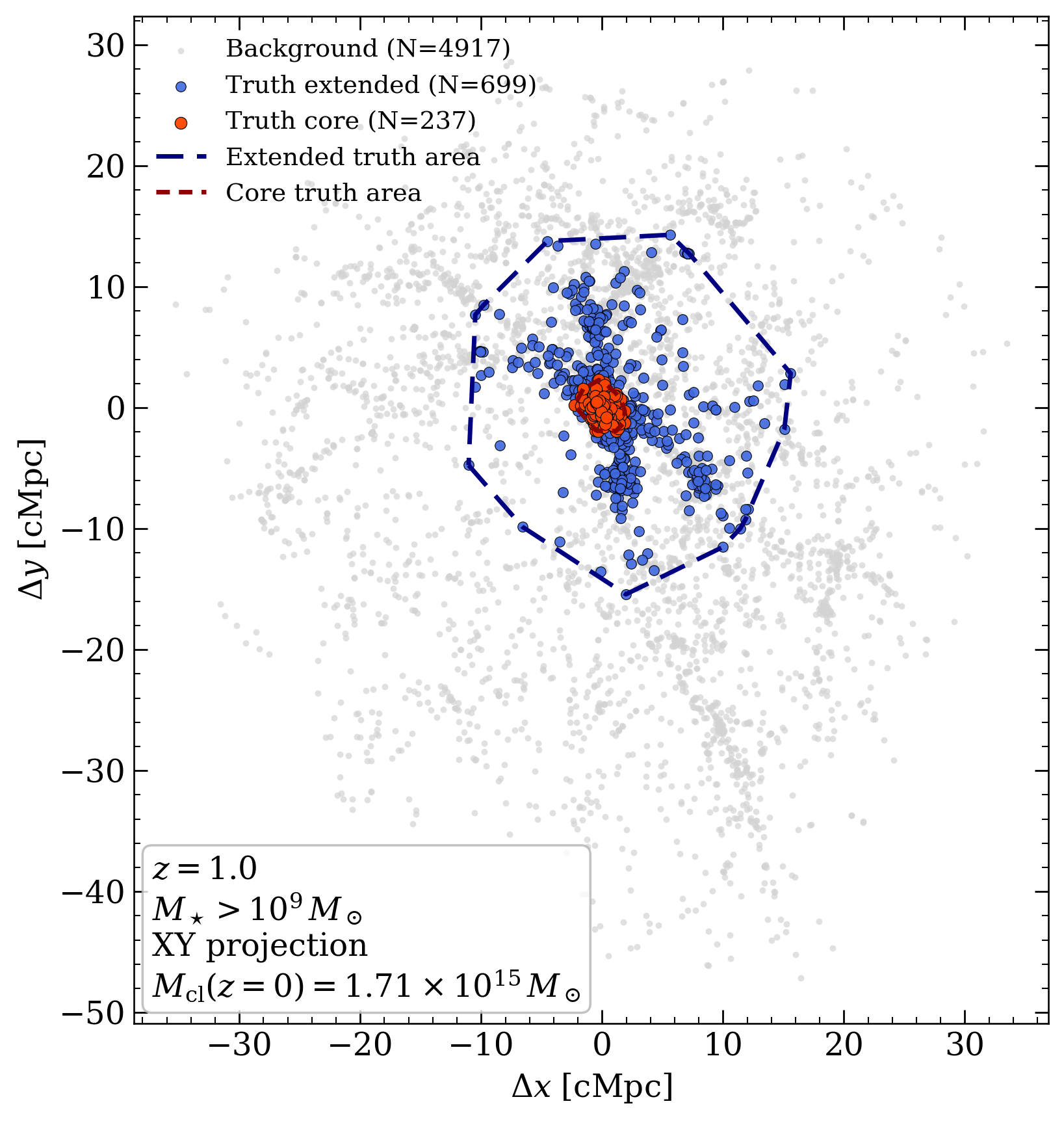}
    \caption{
        Illustrative example of \prodigy\ detections for a representative cluster region at $z=1$, in the $xy$ projection, using galaxies with $M_\star>10^{9}\modot$. The left panel shows the projected galaxy distribution together with the extended protocluster truth set and the core truth set, as well as their corresponding projected convex-hull areas. The middle panel shows the Step~1 detection, including correctly recovered members, missed true members, interlopers, and the projected areas of the extended truth and the Step~1 candidate. The right panel shows the corresponding Step~2 detection relative to the core truth population; the inset provides a zoomed view of the compact central region. This example illustrates the two-level behaviour of \prodigy: Step~1 recovers the extended protocluster environment, while Step~2 identifies a more compact and purer protocluster core.
    }
    \label{Fig:Results:Example}
\end{figure*}

Figure~\ref{Fig:Results:Example} illustrates a representative region at $z=1$ in the $xy$ projection with $M_\star>10^{9}\modot$. It is a visual guide to the two targets, not a summary of the full redshift, projection, mass, or parameter dependence. The extended truth includes a dense central concentration and a diffuse population along surrounding structures. Step~1 recovers $N_{\rm match}=635$ of $N_{\rm truth}=699$ extended members ($C\simeq0.91$), but selects $N_{\rm interlopers}=733$ ($P\simeq0.46$). Its broad footprint therefore achieves high recovery with appreciable contamination.

Step~2 instead concentrates on the core truth, recovering $N_{\rm match}=223$ of $N_{\rm truth}=237$ members ($C\simeq0.94$) with $N_{\rm interlopers}=145$ ($P\simeq0.61$). It reduces the projected extent and increases purity relative to Step~1 while retaining high completeness for the core. The two steps target different physical populations: the extended environment and the robust core. Their footprints also illustrate why member recovery must be considered together with area when defining efficient follow-up regions. The morphology and numerical values vary between realizations; the following subsections quantify the two-level recovery problem statistically.

\subsection{Step 1 performance: recovering extended protocluster regions}
\label{Sec:Results:Step1}
For the representative $z=2$ case with $M_\star>10^{10}\modot$, the Step~1 parameter landscape shows the expected purity--completeness trade-off (Appendix~\ref{App:Step1Landscape}, Fig.~\ref{Fig:Results:Step1Grid}). Low $\alpha$ and large $\ell_1$ connect much of the extended truth population, but include more interlopers and frequently exceed the truth footprint. High $\alpha$ and small $\ell_1$ select cleaner, smaller candidates but miss diffuse components. The largest meanPC values occur at intermediate settings near the allowed--forbidden transition.

This behaviour is similar across the explored nearest-neighbour scales: the optima belong to a family of solutions rather than a single fine-tuned point. The area criterion excludes configurations that gain completeness by selecting an excessively large region. Step~1 therefore recovers a large fraction of the extended population within the parameter range that balances recovery, contamination, and projected area. Appendix~\ref{App:Step1Landscape} gives the full landscape and its interpretation; Appendix~\ref{App:Step1Slices} shows fixed-parameter slices. The redshift and mass-selection dependence of the selected configurations is reported in Sect.~\ref{Sec:Results:Dependencies}.

\subsection{Step 2 performance: identifying robust cores}
\label{Sec:Results:Step2}
For the same $z=2$, $M_\star>10^{10}\modot$ case, we fix the Step~1 parameters to the best allowed extended-region configuration and vary $p_2$. The full response curve is shown in Appendix~\ref{App:Step2Diagnostic}, Fig.~\ref{Fig:Results:Step2ll2p}. At low $p_2$, short adaptive linking lengths give very pure but incomplete cores. Increasing $p_2$ connects more real core members, raising completeness while lowering purity as galaxies outside the core truth set enter the candidate.

MeanPC peaks at intermediate percentiles. The fiducial solution maximises this balance within the allowed area regime, near the transition where the candidate footprint is comparable to, or smaller than, the core footprint. This prevents high completeness from being obtained with an overly extended core selection. Step~2 thus produces a compact candidate with high completeness relative to the core truth population, suitable for prioritising membership confirmation and follow-up. Its systematic redshift and mass-selection dependence is shown below.

\subsection{Dependence on redshift, projection and stellar-mass selection}
\label{Sec:Results:Dependencies}

\begin{figure*}
    \centering
    \includegraphics[width=\textwidth]{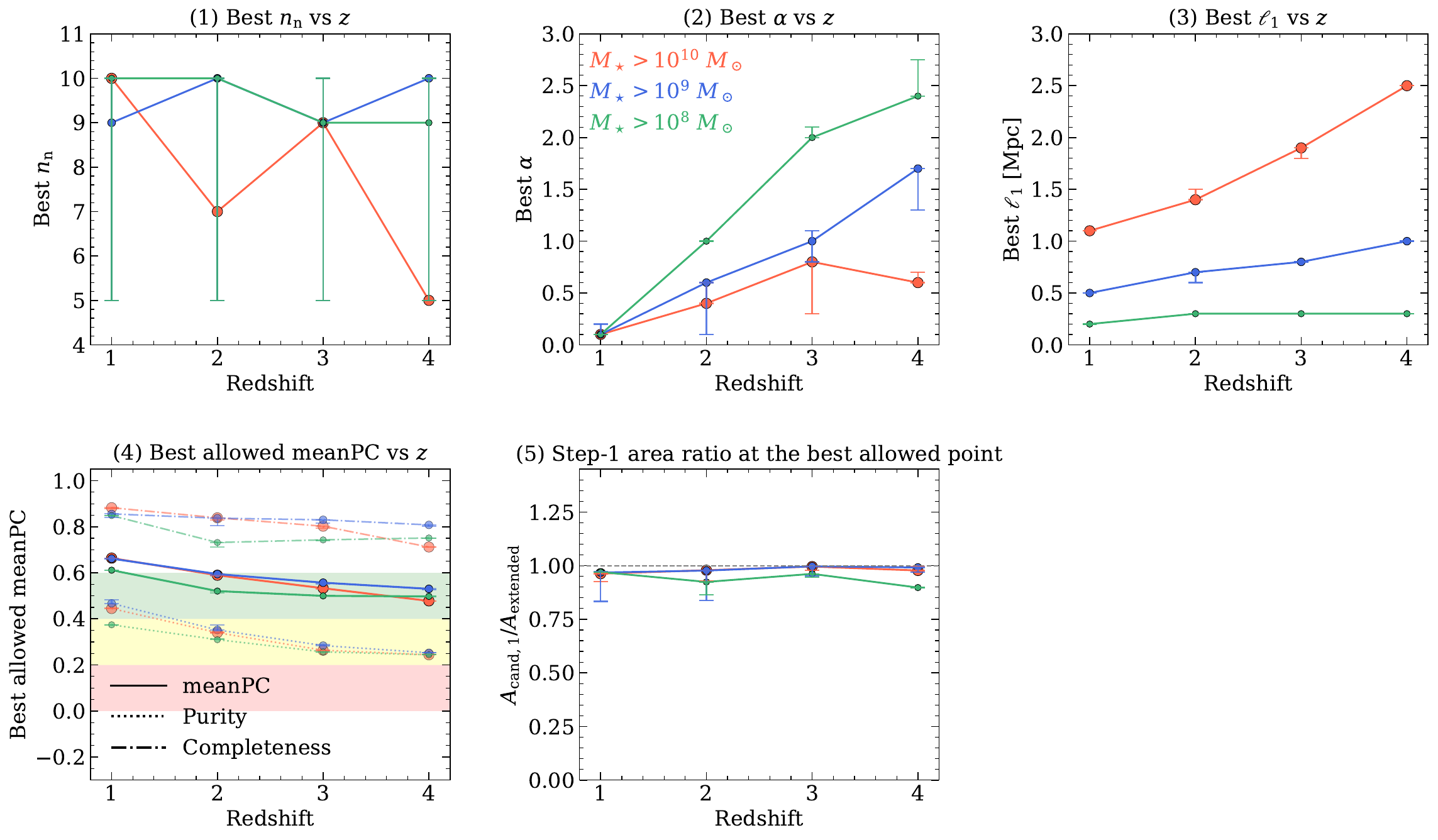}
    \caption{
        Summary of the best allowed Step~1 configurations obtained by maximizing meanPC as a function of redshift, using the three orthogonal projected views. The red, blue, and green curves correspond to the stellar-mass thresholds $M_\star>10^{10}\modot$, $M_\star>10^{9}\modot$, and $M_\star>10^{8}\modot$, respectively. Panels~(1)--(3) show the optimal values of the Step~1 parameters $n_{\rm n}$, $\alpha$, and $\ell_1$, respectively. Panel~(4) shows the maximum allowed meanPC at each redshift, together with the purity and completeness measured at the same meanPC-optimal configuration. The shaded horizontal bands provide a visual guide to low, intermediate, and high meanPC regimes. Panel~(5) shows the area ratio $A_{\rm cand,1}/A_{\rm extended}$ evaluated at the best allowed configuration. Vertical bars indicate the minimum-to-maximum range among the best allowed configurations satisfying $\mathrm{meanPC}\geq\mathrm{meanPC}_{\max}-0.02$, where $\mathrm{meanPC}_{\max}$ is the global maximum for each redshift and stellar-mass threshold. These ranges quantify parameter sensitivity rather than statistical uncertainties. The figure summarizes how the optimal Step~1 performance and parameter choices evolve with redshift and stellar-mass selection.
    }
    \label{Fig:Results:Dependencies:Step1Summary}
\end{figure*}

\begin{figure*}
    \centering
    \includegraphics[width=\textwidth]{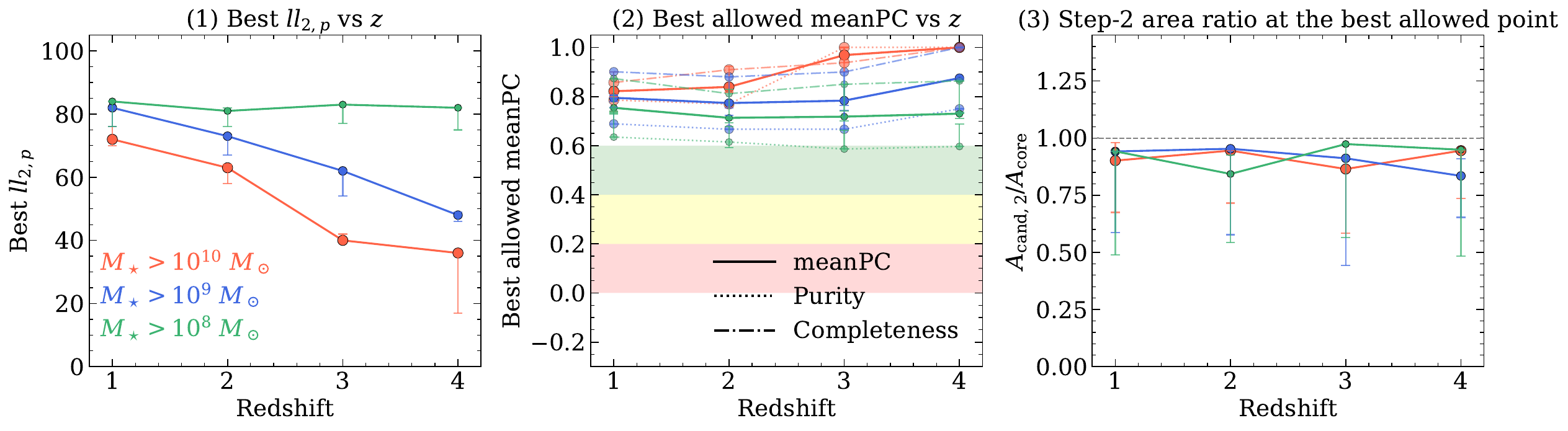}
    \caption{
        Summary of the best allowed Step~2 configurations obtained by maximizing meanPC as a function of redshift, using the three orthogonal projected views and fixing the optimal Step~1 parameters. The red, blue, and green curves correspond to the stellar-mass thresholds $M_\star>10^{10}\modot$, $M_\star>10^{9}\modot$, and $M_\star>10^{8}\modot$, respectively. Panel~(1) shows the optimal value of the Step~2 percentile parameter $p_2$, which defines the adaptive linking length $\ell_2$. Panel~(2) shows the maximum allowed meanPC at each redshift, together with the purity and completeness measured at the same meanPC-optimal configuration. The shaded horizontal bands provide the same visual guide to low, intermediate, and high meanPC regimes as in Fig.~\ref{Fig:Results:Dependencies:Step1Summary}. Panel~(3) shows the Step~2 area ratio, $A_{\rm cand,2}/A_{\rm core}$, evaluated at the best allowed configuration. Vertical bars indicate the minimum-to-maximum range among allowed configurations satisfying $\mathrm{meanPC}\geq\mathrm{meanPC}_{\max}-0.02$, where $\mathrm{meanPC}_{\max}$ is the maximum allowed value for each redshift and stellar-mass threshold. These ranges quantify parameter sensitivity rather than statistical uncertainties. The figure summarizes how the optimal Step~2 core selection evolves with redshift and stellar-mass threshold.
    }
    \label{Fig:Results:Dependencies:Step2Summary}
\end{figure*}

The previous subsections illustrated the behaviour of \prodigy\ through representative cases. We now extend the analysis to the full set of redshifts and stellar-mass selections considered in this work. Rather than showing the complete parameter-space maps for every combination of redshift, projected view, stellar-mass threshold, and algorithmic step, we summarize the results using the best allowed configurations obtained by maximizing meanPC. This provides a compact view of both the performance of \prodigy\ and the parameter values required to achieve that performance.

We first assess the impact of projection. Since protoclusters are intrinsically extended and anisotropic, their appearance may depend on the line of sight. However, we find that the statistical performance of \prodigy\ is only weakly affected by the choice of projected view. Appendix~\ref{App:ProjectionDependence} compares one-dimensional Step~1 diagnostics obtained from individual and combined projections, showing that the purity, completeness, meanPC, and allowed--forbidden transition remain consistent within the region-to-region scatter. For this reason, in the following analysis we combine the three orthogonal projections, increasing the statistical sample size.

For Step~1, Fig.~\ref{Fig:Results:Dependencies:Step1Summary}
summarizes the best allowed configurations obtained by
maximizing meanPC at each redshift and for the three
stellar-mass thresholds, using the three orthogonal projected
views. The figure shows the optimal values of $n_{\rm n}$,
$\alpha$, and $\ell_1$, together with the corresponding
maximum allowed meanPC, purity, completeness, and area
ratio. Vertical bars show the range of these quantities
among near-optimal allowed configurations satisfying
$\mathrm{meanPC}\geq\mathrm{meanPC}_{\max}-0.02$, allowing
us to assess the sensitivity of PRODIGY to its
parameter choices.

Several trends are apparent. The optimal overdensity threshold $\alpha$ increases with redshift, particularly for the lower stellar-mass thresholds, indicating that progressively stronger overdensity cuts are preferred at earlier cosmic times. Similarly, the optimal linking length $\ell_1$ increases towards higher redshifts, especially for the more massive tracer samples, while remaining nearly constant for $M_\star>10^{8}\modot$.
At fixed redshift, $\ell_1$ is systematically larger for more massive tracers, reflecting their lower number densities and larger typical projected separations. The narrow ranges of near-optimal $\ell_1$ values indicate that the preferred connectivity scale is comparatively well constrained within the explored configurations, whereas $\alpha$ exhibits somewhat broader variations.

In contrast, although the optimal $n_{\rm n}$ varies with redshift and stellar-mass selection, the near-optimal configurations span the explored range $5\leq n_{\rm n}\leq10$, with meanPC differing from the global maximum by no more
than 0.02. This demonstrates that Step~1 is only weakly sensitive to the number of nearest neighbours within the tested interval. Consequently, $n_{\rm n}=5$ provides a practical default for applications to large galaxy catalogues, while $\alpha$ and $\ell_1$ should remain calibrated to the redshift and stellar-mass selection.

The performance of Step~1 declines gradually with redshift. The maximum allowed meanPC decreases from $z=1$ to $z=4$ for all three stellar-mass thresholds, primarily driven by a reduction in purity. Across the three stellar-mass selections, completeness (purity) reaches approximately 90\% (50\%) at $z=1$, decreasing to 70\% (25\%) at $z=4$. Thus, Step~1 continues to recover a substantial fraction of the extended protocluster population at high redshift, but with increasing contamination. The three stellar-mass selections yield broadly comparable maximum allowed meanPC values, indicating a similar balance between purity and completeness across the explored tracer samples.

The optimal area ratios remain close to unity across the three stellar-mass selections, with relatively narrow ranges among near-optimal configurations. This indicates that the recovered projected areas are comparable to those of the true extended protocluster populations and show limited sensitivity to parameter choices within the near-optimal regime.

\begin{figure*}
    \centering
    \includegraphics[width=\textwidth]{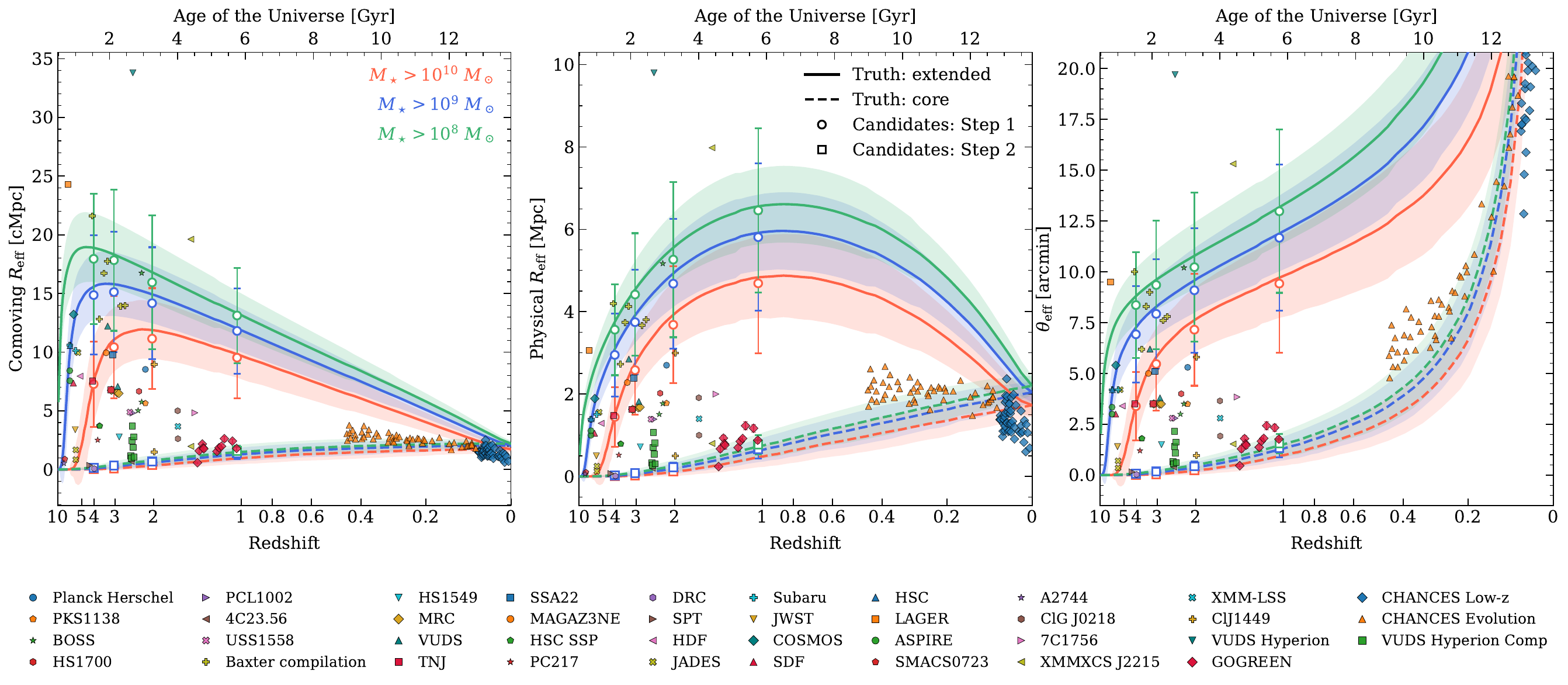}
    \caption{Effective radii of the descendant-defined truth populations and \prodigy\ candidates, compared with observational footprint proxies in comoving (left), physical (middle), and angular (right) units. Solid and dashed curves denote extended and core truth; open circles and squares denote the best allowed meanPC Step~1 and Step~2 candidates. Red, blue, and green denote $M_\star>10^{10}$, $10^{9}$, and $10^{8}\modot$, respectively. Bands and candidate error bars indicate scatter across regions and projections. Additional symbols identify the observational samples listed in the legend. Their heterogeneous apertures, field areas, projected extents, and $R_{200}$ values are footprint proxies rather than a uniform radius definition; CHANCES and GOGREEN are cluster-scale anchors. The source mapping and proxy conventions are given in Appendix~\ref{App:ObservationalFootprints}. Candidate markers cover the calibrated redshifts $z=1$--4; curves outside that interval do not establish detection performance there.}
    \label{Fig:Results:BestParameters:Reff}
\end{figure*}

The corresponding Step~2 results are shown in Fig.~\ref{Fig:Results:Dependencies:Step2Summary}, where $p_2$ is optimized against the core truth population with the Step~1 parameters fixed at their best allowed values. The optimal Step~2 percentile $p_2$ decreases with redshift for the $M_\star>10^{10}\modot$ and $M_\star>10^{9}\modot$ samples, whereas it remains nearly constant at $p_2\sim80$ for $M_\star>10^{8}\modot$. Thus, the more massive tracer selections favour progressively lower percentiles of the internal galaxy-separation distribution at earlier epochs, while the lowest-mass selection shows little redshift dependence. This behaviour is consistent with the purpose of Step~2: once the extended overdensity has been identified, the algorithm must isolate the densest and most reliable subset of that structure rather than follow its full filamentary extent. At fixed redshift, lower stellar-mass thresholds generally favour higher $p_2$, consistent with their denser sampling of the protocluster region. The relatively narrow ranges among near-optimal configurations indicate that the preferred percentile is generally well constrained, although broader intervals occur in some cases.

Step~2 maintains high performance across the explored redshift range, although its evolution depends on stellar-mass selection. The maximum allowed meanPC increases towards higher redshifts for $M_\star>10^{10}\modot$, approaching unity at $z=4$, and reaches approximately 0.9 for $M_\star>10^{9}\modot$. In contrast, it remains relatively stable at $\sim0.7$--0.8 for $M_\star>10^{8}\modot$. Across the three selections, completeness (purity) ranges from approximately 85--90\% (65--80\%) at $z=1$ to 85--100\% (60--100\%) at $z=4$. Thus, the values of purity and completeness measured at the meanPC-optimal point show that Step~2 is able to identify a compact core population with a favourable balance between reliability and recovery.

The area ratios shown in the right panel remain below unity for all redshifts and all three stellar-mass thresholds. This confirms that the best allowed Step~2 configurations do not identify the core by selecting regions larger than the projected core footprint. Instead, the optimal core candidates occupy areas comparable to, but slightly smaller than, the true core area. This is precisely the desired behaviour for a robust core-finding stage: the detection remains compact while preserving high membership-based performance.

\subsection{Best configurations and recommended parameter choices}
\label{Sec:Results:BestParameters}
We adopt the best allowed meanPC configurations as the fiducial calibration of \prodigy. Table~\ref{Tab:Results:BestParameters} lists the optimal Step~1 parameters ($n_{\rm n}$, $\alpha$, and $\ell_1$) and the corresponding Step~2 percentile $p_2$ for each redshift and stellar-mass threshold. Alternative configurations maximizing purity or completeness may be adopted for more specialised selection goals.

\begin{table}
    \caption{Fiducial \prodigy\ parameters obtained by maximizing the allowed meanPC for each redshift and stellar-mass threshold. Superscripts and subscripts indicate the upper and lower deviations from the optimal values among near-optimal allowed configurations satisfying $\mathrm{meanPC}\geq\mathrm{meanPC}_{\max}-0.02$. Zero-width intervals are omitted. These ranges represent parameter sensitivity, not statistical uncertainties.}
    \label{Tab:Results:BestParameters}
    \centering
    \normalsize
    \setlength{\tabcolsep}{10pt}
    \begin{tabular}{cccccc}
        \hline\hline
        $z$ & $M_{\star,\min}$ & $n_{\rm n}$ & $\alpha$ & $\ell_1$ & $p_2$ \\
        & ($M_\odot$) & & & (cMpc) & \\
        \hline
        1 & $10^{10}$ & $10^{+0}_{-5}$ & $0.1$ & $1.1$ & $72^{+4}_{-2}$ \\
        1 & $10^{9}$  & $9^{+1}_{-4}$  & $0.1^{+0.1}_{-0}$ & $0.5$ & $82^{+0}_{-6}$ \\
        1 & $10^{8}$  & $10^{+0}_{-5}$ & $0.1$ & $0.2$ & $84^{+0}_{-8}$ \\
        \hline
        2 & $10^{10}$ & $7^{+3}_{-2}$ & $0.4^{+0.2}_{-0}$ & $1.4^{+0.1}_{-0}$ & $63^{+1}_{-5}$ \\
        2 & $10^{9}$  & $10^{+0}_{-5}$ & $0.6^{+0}_{-0.5}$ & $0.7^{+0}_{-0.1}$ & $73^{+0}_{-6}$ \\
        2 & $10^{8}$  & $10^{+0}_{-5}$ & $1.0$ & $0.3$ & $81^{+1}_{-5}$ \\
        \hline
        3 & $10^{10}$ & $9^{+1}_{-4}$ & $0.8^{+0}_{-0.5}$ & $1.9^{+0}_{-0.1}$ & $40^{+2}_{-0}$ \\
        3 & $10^{9}$  & $9^{+1}_{-4}$ & $1.0^{+0.1}_{-0.2}$ & $0.8$ & $62^{+0}_{-8}$ \\
        3 & $10^{8}$  & $9^{+1}_{-4}$ & $2.0^{+0.1}_{-0}$ & $0.3$ & $83^{+0}_{-6}$ \\
        \hline
        4 & $10^{10}$ & $5^{+5}_{-0}$ & $0.6^{+0.1}_{-0}$ & $2.5$ & $36^{+0}_{-19}$ \\
        4 & $10^{9}$  & $10^{+0}_{-5}$ & $1.7^{+0}_{-0.4}$ & $1.0$ & $48^{+1}_{-2}$ \\
        4 & $10^{8}$  & $9^{+1}_{-4}$ & $2.4^{+0.35}_{-0}$ & $0.3$ & $82^{+0}_{-7}$ \\
        \hline
    \end{tabular}
\end{table}

The broad near-optimal ranges of $n_{\rm n}$ in Fig.~\ref{Fig:Results:Dependencies:Step1Summary} indicate that Step~1 performance is relatively insensitive to the number of nearest neighbours within $5\leq n_{\rm n}\leq10$. We therefore recommend $n_{\rm n}=5$ as a practical alternative for large galaxy catalogues, potentially reducing the computational cost of local-density estimation. This choice requires the corresponding calibration of $\alpha$ and $\ell_1$, as well as reassessing the Step~2 optimum. Its computational advantage remains to be quantified.

Candidate areas near the truth footprints suggest that the calibration can define efficient follow-up regions, without overly compact selections that miss members or overly broad selections dominated by contaminants. For a more direct size interpretation, we define the effective radius as $R_{\rm eff}=\sqrt{A/\pi}$ and express the radii in comoving, physical, and angular units (see Fig.~\ref{Fig:Results:BestParameters:Reff}). For the three mass thresholds shown in this figure and both steps, candidate radii closely track the corresponding truth populations: Step~1 follows the extended size evolution and Step~2 the smaller core scale. The agreement persists in physical and angular units, which are directly relevant to observations. The accompanying literature points are examined separately in Sect.~\ref{Sec:Results:ObservedFootprints}.

This result strengthens the interpretation of the area-ratio analysis presented above. The fact that the best allowed meanPC configurations recover not only similar projected areas but also similar effective radii implies that \prodigy\ can provide realistic estimates of the characteristic size of both the extended protocluster region and its compact core. In an observational context, this means that the algorithm can be used not only to identify candidate protoclusters, but also to define physically motivated scales for follow-up strategies. The Step~1 candidates provide a calibrated estimate of the projected extent over which the broader protocluster environment is expected, whereas the Step~2 candidates identify a much more compact region that is well suited for conservative spectroscopic targeting.

\subsection{Observed footprints: extended environments and compact cores}
\label{Sec:Results:ObservedFootprints}
Figure~\ref{Fig:Results:BestParameters:Reff} extends the size comparison to published observational footprint proxies. These are the sizes of the selected or surveyed regions, not a homogeneous sample of intrinsic protocluster radii. The compilation combines aperture and membership radii, field-of-view equivalents, projected extents, volume-derived proxies, and cluster $R_{200}$ measurements. Appendix~\ref{App:ObservationalFootprints} maps the legend groups to their sources and records the distinctions required to interpret them. We use the supplied compilation for a comparison of spatial scales, without treating its entries as a statistical fit to the simulated populations.

The comoving panel gives the clearest separation between the two simulated targets. At the calibrated redshifts, extended truth and Step~1 candidates occupy scales of approximately $10$--$20$ cMpc, depending on redshift and tracer selection, whereas core truth and Step~2 candidates occupy much smaller scales, typically below a few cMpc. Observational footprints span both regimes: SSA22, the broad BOSS and VUDS selections, and several Subaru and other high-redshift overdensities sample extended-environment scales, while the selected DRC, SPT2349-56, and JADES-ID1 footprints are compact. These are comparisons with the particular footprints plotted, rather than exclusive classifications of the underlying systems: the same object can have a compact detected component and a much larger surrounding environment. The high-redshift points also contextualise spatial scales beyond $z=4$, but do not validate \prodigy\ outside its calibration interval.

Hyperion illustrates this hierarchy within a single observed structure \citep{cucciati2018_hyperion}. Its full projected footprint, represented by the area-equivalent radius of the supplied $60\times60$ cMpc$^2$ region, lies above the extended curves, consistent with a proto-supercluster containing multiple components. The seven component proxies instead span approximately $0.97$--$3.72$ cMpc, approaching the compact regime. A large observational footprint can thus contain several dense peaks; neither the total footprint nor an individual peak should automatically be equated with the progenitor population of one final cluster.

GOGREEN provides a complementary comparison at $z\sim1$--1.4 \citep{old2020_gogreen}. Its plotted $R_{200\mathrm{c}}$ values lie near the core or halo scale, well below the extended progenitor footprints. They anchor already concentrated cluster regions rather than measure the full environment of their descendant-defined members. CHANCES supplies the analogous low-redshift anchor \citep{sifon2025_chances}: the plotted values are $R_{200}$, not the survey coverage extending to several virial radii. The large low-redshift angular radii mainly reflect distance geometry. Neither sample is a direct measurement of extended high-redshift protocluster size.

The physical panel places compact cores and cluster regions predominantly on sub-Mpc to roughly Mpc scales, while the simulated extended footprints reach several physical Mpc. The broad XMMXCS~J2215 environment and Hyperion lie above typical single-cluster footprints; in the former case the point represents the observed field, whereas the separately plotted core uses $R_{200}$ \citep{hayashi2011}. The Cl~J1449 central and environmental apertures similarly illustrate why one object can have more than one reported scale \citep{coogan2018,smith2019}. A physical aperture of order $1$ Mpc at $z\sim1$--2 therefore need not contain the complete progenitor environment.

The angular panel translates the distinction into follow-up requirements. At $z\sim2$--4, extended Step~1-like regions span several to roughly $5$--$15$ arcmin in effective radius, while compact cores generally require much smaller fields. A small-field JWST detection can identify a real overdensity while sampling only a dense component. Conversely, mosaics and wider surveys can map a larger footprint, as illustrated by the explicit ASPIRE mosaic \citep{champagne2025}. A compact observed footprint should consequently not be taken as evidence for an intrinsically small full protocluster.

Finally, the three stellar-mass selections are not interchangeable. The $M_\star>10^{8}\modot$ population traces larger extended footprints than the more massive selections in this comparison, consistent with additional sampling of outskirts and lower-density components. This motivates accounting for tracer depth when comparing survey footprints with the simulation, but does not establish a universal mapping between a stellar-mass cut and an LAE, SMG, or flux-selected sample. Taken together, the panels show that published ``protocluster sizes'' mix extended environments, compact components, and cluster-scale regions. The simulation supplies a physical reference for interpreting these observational selections while retaining their different definitions.

\section{Discussion}
\label{Sec:Discussion}

The results presented above show that protocluster detection is not a single-scale problem. The same underlying structure can be described either as an extended progenitor region, containing a large fraction of the galaxies that will become part of the final cluster, or as a compact core, containing the densest and most reliable subset of that population. In this section, we discuss the implications of this two-level view, the relation of \prodigy\ to previous protocluster-finding approaches, its relevance for observational surveys, and the main limitations of the present calibration.

\subsection{Why the two-step structure matters}
\label{Sec:Discussion:TwoStep}
The two stages answer different questions: where the extended progenitor environment lies in projection, and which part forms its compact, robust core. They require distinct targets and definitions of success.

The need for this distinction follows directly from the physical nature of protoclusters. At high redshift, the progenitors of present-day clusters are not virialized systems with a well-defined centre and radius. They are extended, anisotropic, and often fragmented structures, with true members distributed across several haloes, groups, and filamentary components. A method optimized only to find the densest region will therefore miss a substantial fraction of the final cluster progenitor population. Conversely, a method optimized only to recover as many true members as possible can select projected regions that are too extended and contaminated to be useful for observational follow-up.

Step~1 connects overdense galaxies over larger scales, especially at high redshift and for massive tracers, and recovers the extended population with high completeness but appreciable interlopers. The area condition limits this expansion. Step~2 adapts to the internal spacing and isolates a smaller, higher-confidence core for follow-up. The opposite parameter trends---increasing $\ell_1$ but decreasing $p_2$ with redshift---show that the method calibrates distinct components rather than simply repeats a FoF search.

Purity and completeness must accordingly be interpreted relative to each target. Moderate purity can accompany a useful Step~1 recovery if completeness is high and the area ratio is near unity; purity is more central for the compact Step~2 target. The two outputs allow the selection strategy to follow the scientific goal, from mapping the full environment or descendant population to efficient spectroscopic confirmation.

\subsection{Comparison with previous protocluster-finding approaches}
\label{Sec:Discussion:LiteratureComparison}
Tracer-based overdensity searches have discovered candidates across a wide redshift range using LAEs, LBGs, massive galaxies, SMGs, quasars, radio galaxies, and photometric-redshift selections \citep[e.g.][]{castignani2014,chiang2014,chiang2015,toshikawa2018,calvi2023,brinch2023,zhang2026}. Their tracer, aperture, smoothing, and redshift choices do not necessarily measure the complete descendant population. Simulations have related overdensities to protocluster probability and descendant mass \citep{chiang2013,chiang2014}, established the extent and membership beyond the main progenitor \citep{muldrew2015,muldrew2018}, and quantified aperture- and geometry-dependent purity and completeness \citep{lovell2018}.

\prodigy\ builds on this framework with descendant-defined member accounting, separate extended and core targets, and an area calibration against the corresponding truth footprints. A single threshold or fixed aperture cannot generally optimise both targets. The allowed condition rejects configurations that gain completeness by overextending the candidate, balancing member recovery, contamination, and size. \prodigy\ therefore complements observational overdensity searches with simulation-based interpretation and selection optimisation, including physically motivated scales for follow-up of large survey samples.

\subsection{Observational implications}
\label{Sec:Discussion:Observations}
The observational comparison in Sect.~\ref{Sec:Results:ObservedFootprints} makes the two-level interpretation concrete. In the simulations, an extended protocluster is the selected galaxy population linked to a common $z=0$ cluster, whereas its core is the descendant-linked subset already associated with the main progenitor system. These definitions do not depend on an observational aperture. Observations instead identify overdensities within a selected footprint and cannot follow individual galaxies to their future descendants. \prodigy\ therefore provides a reference for asking whether an observed selection samples extended-environment or compact-core scales. Size agreement alone does not prove common descendant membership or identify an observed density peak as the unique main-progenitor core.

Beyond protocluster identification, the two-scale structure of PRODIGY may provide a basis for statistical predictions of descendant cluster mass. While the compact Step~2 region traces the main progenitor system, the extended Step~1 footprint samples a larger fraction of the environment that will eventually assemble into the $z=0$ cluster. Its size, richness, and galaxy overdensity may therefore contain complementary information about the final halo mass, as suggested by previous simulation-based calibrations \citep{chiang2013,chiang2014,lovell2018}. This prospect is also motivated by \citet{Witten2026}, who highlight the potential importance of considering halo mass over Lagrangian scales when linking high-redshift protoclusters to their present-day descendants. However, projected effective radii alone do not provide a direct mass estimator, and the present calibration is restricted to progenitors of massive present-day clusters. Establishing a statistical relation between PRODIGY observables and $M_{200c}(z=0)$, including its intrinsic scatter and extension to lower-mass descendants, is therefore a promising direction for future work.

This distinction matters for Hyperion's broad structure and its compact components, for GOGREEN and CHANCES cluster-scale anchors, and for small-field high-redshift detections. A core-like observational selection can be embedded in a much larger progenitor environment; finding a compact component and recovering the complete protocluster are different observing goals. Step~1 motivates mapping the surrounding environment, whereas Step~2 prioritises dense regions for spectroscopic confirmation or detailed multi-wavelength work. The angular panel provides a practical guide to the field coverage needed for each goal, conditional on tracer depth and redshift.

Euclid, LSST, HSC, PFS, and JWST programmes \citep{laureijs2011,ivezic2019,aihara2018,takada2014,gardner2006} offer complementary coverage and tracer selections. An observational implementation should match the redshift slices, tracer density, mask, completeness, and selection function to survey-specific mocks before transferring the calibration. The present footprint comparison motivates this workflow; it is not a measurement of observational purity or completeness, nor a deterministic conversion from any detected core to its full descendant population.

\subsection{Limitations and future work}
\label{Sec:Discussion:Limitations}
Our merger-tree truth is physically defined but inaccessible for individual observed candidates; the reported performance is a simulation-calibrated expectation, not a directly measurable observational quantity. Stellar-mass selections also differ from real samples limited by apparent magnitude, colour, emission-line detectability, redshift quality, or spectroscopic completeness. Different tracers may require different parameters, so mass-threshold tests do not replace survey-specific calibration.

The literature overlay also spans heterogeneous cosmologies, geometric proxies, and redshifts beyond our $z=1$--4 calibration. Lens-corrected source-plane radii must be distinguished from image-plane field sizes, and a volume-derived proxy must not be interpreted as a directly measured projected radius. These differences preclude a homogeneous size-distribution test or definitive object-by-object descendant classification from this figure alone.

The projected catalogues are not full mock observations. Three orthogonal views quantify viewing-direction dependence but omit lightcone geometry, redshift-space distortions, photometric-redshift errors, survey masks, variable depth, fibre collisions, and realistic foreground and background contamination. These effects matter particularly when photometric line-of-sight uncertainties exceed protocluster sizes, and spectroscopic recovery depends on the adopted velocity window.

Photometric extensions should use mocks selected by apparent magnitude, colour, and redshift probability distributions, with extended detections in slices or tomographic bins and cores for follow-up. Spectroscopic extensions can add redshift or velocity cuts relative to the candidate centre to test line-of-sight coherence. Future work will consider semi-analytic and dark-matter-only catalogues, lightcones, and survey-specific mocks, enabling tests of lower stellar masses, other tracers, survey depth, and redshift precision. The current work supplies the baseline calibration; \prodigy~2.0 will explicitly include photometric and spectroscopic selection functions before direct survey applications and catalogues with extended and core classifications.

\section{Summary and conclusions}
\label{Sec:SummaryConclusions}
In this work, we have introduced \prodigy, a two-step algorithm designed to identify the high-redshift progenitors of present-day galaxy clusters in projected galaxy distributions. The method was calibrated using the 7K-GIZMO hydrodynamical galaxy catalogues of {\sc The Three Hundred} Project, and evaluated against physically defined protocluster ground-truth populations obtained by tracing galaxies to their descendant clusters at $z=0$. Our main conclusions are as follows.

\begin{enumerate}
    \item \prodigy\ identifies two distinct protocluster populations calibrated against descendant-defined ground truths: Step~1 recovers the extended progenitor environment, while Step~2 identifies the compact core associated with the main progenitor system.

    \item Step~1 achieves high completeness with appreciable contamination. Across the three stellar-mass selections, completeness (purity) decreases from approximately 90\% (50\%) at $z=1$ to 70\% (25\%) at $z=4$, with broadly comparable meanPC values across the explored mass thresholds.

    \item Step~2 recovers compact cores with completeness (purity) of approximately 85--90\% (65--80\%) at $z=1$ and 85--100\% (60--100\%) at $z=4$. Unlike Step~1, its performance shows a stronger dependence on stellar-mass selection.

    \item The optimal parameters depend on redshift and tracer selection. Step~1 linking lengths generally increase with redshift and stellar-mass threshold, while $p_2$ decreases with redshift for the two higher-mass selections but remains nearly constant for $M_\star>10^{8}\modot$. The weak sensitivity to $n_{\rm n}$ within $5\leq n_{\rm n}\leq10$ provides flexibility in parameter selection. Table~\ref{Tab:Results:BestParameters} provides the calibrated configurations.

    \item The area-ratio criterion prevents excessive footprint growth. At the best allowed meanPC configurations, candidate areas remain close to those of their corresponding truth populations. Their effective radii reproduce the distinct scales of extended protocluster environments and compact cores.

    \item Comparison with observational footprints highlights the distinction between extended environments, compact components, and virialized cluster scales. These comparisons provide physically motivated scales for observational follow-up, although individual classifications remain dependent on tracer selection, geometry, and survey-specific calibration.

\end{enumerate}

\prodigy\ provides a simulation-calibrated bridge between observed galaxy overdensities and the physically defined progenitors of present-day clusters, offering a framework for interpreting protocluster candidates in simulations and mock surveys.
Future work will investigate whether Step~1 sizes, richness, and overdensities can statistically constrain descendant halo masses at $z=0$.
Further extensions incorporating photometric and spectroscopic selection functions, redshift uncertainties, and lightcone geometry will enable its application to real observational surveys.

\begin{acknowledgements}
    This work has been made possible by {\sc The Three Hundred} (https://the300-project.org) collaboration.
    JSG acknowledges funding from Predoctoral contract `Formaci\'on de Personal Investigador' from the Universidad Aut\'onoma de Madrid (FPI-UAM, 2021).
    G.Y., JSG and D.C. %
    would like to thank the Spanish Agencia Estatal de Investigación of the  Ministerio de Ciencia e Innovaci\'on (AEI/MICINN) for financial support under the project grant PID2024-156100NB-C21.
    D.C. is further supported by CNS2024-154550, funded by MCIN/AEI/10.13039/501100011033.
    TH acknowledge Consejo Nacional de Investigaciones Cient\'ificas y T\'ecnicas (CONICET), Argentina, for their supporting fellowships.
    The authors acknowledge The Red Espa\~nola de Supercomputaci\'on for granting computing time for running the hydrodynamical %
    simulations of {\sc The Three Hundred} galaxy cluster project in the Marenostrum supercomputer at the Barcelona Supercomputing Center and Cibeles Supercomputers through various of RES grants.
    The 7K %
    hydrodynamic simulations were performed also on the DIaL3 -- DiRAC Data Intensive service at the University of Leicester through the RAC15 grant: Seedcorn/ACTP317, and on the Niagara supercomputer at the SciNet HPC Consortium. DIaL3 is managed by the University of Leicester Research Computing Service on behalf of the STFC DiRAC HPC Facility (www.dirac.ac.uk).
    The DiRAC service at Leicester was funded by BEIS (ST/K000373/1), UKRI, STFC capital funding, and STFC operations grants (ST/K0003259/1). DiRAC is part of the UKRI Digital Research Infrastructure. SciNet \citep{Loken_2010} is funded by Innovation, Science and Economic Development Canada; the Digital Research Alliance of Canada; the Ontario Research Fund: Research Excellence; and the University of Toronto.
    This work made also use of the local computer facilities at the Universidad Aut\'onoma de Madrid.
\end{acknowledgements}

\section*{Data Availability}
The 7K-GIZMO galaxy catalogues of \textsc{The Three Hundred} Project used in this work are available upon request following the collaboration's data-sharing guidelines, at \url{https://the300.ft.uam.es}. The \prodigy\ source code and documentation are publicly available at \url{https://github.com/JonathanSGomez/prodigy}.

\bibliography{references}

@article{cui2018:the300,
    author = {Cui, Weiguang and Knebe, Alexander and Yepes, Gustavo and Pearce, Frazer and Power, Chris and Dave, Romeel and Arth, Alexander and Borgani, Stefano and Dolag, Klaus and Elahi, Pascal and Mostoghiu, Robert and Murante, Giuseppe and Rasia, Elena and Stoppacher, Doris and Vega-Ferrero, Jesus and Wang, Yang and Yang, Xiaohu and Benson, Andrew and Cora, Sofía A and Croton, Darren J and Sinha, Manodeep and Stevens, Adam R H and Vega-Martínez, Cristian A and Arthur, Jake and Baldi, Anna S and Cañas, Rodrigo and Cialone, Giammarco and Cunnama, Daniel and De Petris, Marco and Durando, Giacomo and Ettori, Stefano and Gottlöber, Stefan and Nuza, Sebastián E and Old, Lyndsay J and Pilipenko, Sergey and Sorce, Jenny G and Welker, Charlotte},
    title = "{The Three Hundred project: a large catalogue of theoretically modelled galaxy clusters for cosmological and astrophysical applications}",
    journal = {Monthly Notices of the Royal Astronomical Society},
    volume = {480},
    number = {3},
    pages = {2898-2915},
    year = {2018},
    month = {08},
    issn = {0035-8711},
    doi = {10.1093/mnras/sty2111},
    url = {https://doi.org/10.1093/mnras/sty2111},
    eprint = {https://academic.oup.com/mnras/article-pdf/480/3/2898/25500565/sty2111.pdf},
}

@article{cui2022:gizmo,
    author = {Cui, Weiguang and Dave, Romeel and Knebe, Alexander and Rasia, Elena and Gray, Meghan and Pearce, Frazer and Power, Chris and Yepes, Gustavo and Anbajagane, Dhayaa and Ceverino, Daniel and Contreras-Santos, Ana and de Andres, Daniel and De Petris, Marco and Ettori, Stefano and Haggar, Roan and Li, Qingyang and Wang, Yang and Yang, Xiaohu and Borgani, Stefano and Dolag, Klaus and Zu, Ying and Kuchner, Ulrike and Cañas, Rodrigo and Ferragamo, Antonio and Gianfagna, Giulia},
    title = "{The Three Hundred project: The gizmo-simba run}",
    journal = {Monthly Notices of the Royal Astronomical Society},
    volume = {514},
    number = {1},
    pages = {977-996},
    year = {2022},
    month = {05},
    issn = {0035-8711},
    doi = {10.1093/mnras/stac1402},
    url = {https://doi.org/10.1093/mnras/stac1402},
    eprint = {https://academic.oup.com/mnras/article-pdf/514/1/977/43984919/stac1402.pdf},
}

@article{klypin2016,
    author = {Klypin, Anatoly and Yepes, Gustavo and Gottlöber, Stefan and Prada, Francisco and Heß, Steffen},
    title = "{MultiDark simulations: the story of dark matter halo concentrations and density profiles}",
    journal = {Monthly Notices of the Royal Astronomical Society},
    volume = {457},
    number = {4},
    pages = {4340-4359},
    year = {2016},
    month = {02},
    issn = {0035-8711},
    doi = {10.1093/mnras/stw248},
    url = {https://doi.org/10.1093/mnras/stw248},
    eprint = {https://academic.oup.com/mnras/article-pdf/457/4/4340/18515365/stw248.pdf},
}

@article{planckcollaboration2016,
	author = {{Planck Collaboration} and {Ade, P. A. R.} and {Aghanim, N.} and {Arnaud, M.} and {Ashdown, M.} and {Aumont, J.} and {Baccigalupi, C.} and {Banday, A. J.} and {Barreiro, R. B.} and {Bartlett, J. G.} and {Bartolo, N.} and {Battaner, E.} and {Battye, R.} and {Benabed, K.} and {Beno\^{\i}t, A.} and {Benoit-L\'evy, A.} and {Bernard, J.-P.} and {Bersanelli, M.} and {Bielewicz, P.} and {Bock, J. J.} and {Bonaldi, A.} and {Bonavera, L.} and {Bond, J. R.} and {Borrill, J.} and {Bouchet, F. R.} and {Boulanger, F.} and {Bucher, M.} and {Burigana, C.} and {Butler, R. C.} and {Calabrese, E.} and {Cardoso, J.-F.} and {Catalano, A.} and {Challinor, A.} and {Chamballu, A.} and {Chary, R.-R.} and {Chiang, H. C.} and {Chluba, J.} and {Christensen, P. R.} and {Church, S.} and {Clements, D. L.} and {Colombi, S.} and {Colombo, L. P. L.} and {Combet, C.} and {Coulais, A.} and {Crill, B. P.} and {Curto, A.} and {Cuttaia, F.} and {Danese, L.} and {Davies, R. D.} and {Davis, R. J.} and {de Bernardis, P.} and {de Rosa, A.} and {de Zotti, G.} and {Delabrouille, J.} and {D\'esert, F.-X.} and {Di Valentino, E.} and {Dickinson, C.} and {Diego, J. M.} and {Dolag, K.} and {Dole, H.} and {Donzelli, S.} and {Dor\'e, O.} and {Douspis, M.} and {Ducout, A.} and {Dunkley, J.} and {Dupac, X.} and {Efstathiou, G.} and {Elsner, F.} and {En\ss{}lin, T. A.} and {Eriksen, H. K.} and {Farhang, M.} and {Fergusson, J.} and {Finelli, F.} and {Forni, O.} and {Frailis, M.} and {Fraisse, A. A.} and {Franceschi, E.} and {Frejsel, A.} and {Galeotta, S.} and {Galli, S.} and {Ganga, K.} and {Gauthier, C.} and {Gerbino, M.} and {Ghosh, T.} and {Giard, M.} and {Giraud-H\'eraud, Y.} and {Giusarma, E.} and {Gjerl\o{}w, E.} and {Gonz\'alez-Nuevo, J.} and {G\'orski, K. M.} and {Gratton, S.} and {Gregorio, A.} and {Gruppuso, A.} and {Gudmundsson, J. E.} and {Hamann, J.} and {Hansen, F. K.} and {Hanson, D.} and {Harrison, D. L.} and {Helou, G.} and {Henrot-Versill\'e, S.} and {Hern\'andez-Monteagudo, C.} and {Herranz, D.} and {Hildebrandt, S. R.} and {Hivon, E.} and {Hobson, M.} and {Holmes, W. A.} and {Hornstrup, A.} and {Hovest, W.} and {Huang, Z.} and {Huffenberger, K. M.} and {Hurier, G.} and {Jaffe, A. H.} and {Jaffe, T. R.} and {Jones, W. C.} and {Juvela, M.} and {Keih\"anen, E.} and {Keskitalo, R.} and {Kisner, T. S.} and {Kneissl, R.} and {Knoche, J.} and {Knox, L.} and {Kunz, M.} and {Kurki-Suonio, H.} and {Lagache, G.} and {L\"ahteenm\"aki, A.} and {Lamarre, J.-M.} and {Lasenby, A.} and {Lattanzi, M.} and {Lawrence, C. R.} and {Leahy, J. P.} and {Leonardi, R.} and {Lesgourgues, J.} and {Levrier, F.} and {Lewis, A.} and {Liguori, M.} and {Lilje, P. B.} and {Linden-V\o{}rnle, M.} and {L\'opez-Caniego, M.} and {Lubin, P. M.} and {Mac\'{\i}as-P\'erez, J. F.} and {Maggio, G.} and {Maino, D.} and {Mandolesi, N.} and {Mangilli, A.} and {Marchini, A.} and {Maris, M.} and {Martin, P. G.} and {Martinelli, M.} and {Mart\'{\i}nez-Gonz\'alez, E.} and {Masi, S.} and {Matarrese, S.} and {McGehee, P.} and {Meinhold, P. R.} and {Melchiorri, A.} and {Melin, J.-B.} and {Mendes, L.} and {Mennella, A.} and {Migliaccio, M.} and {Millea, M.} and {Mitra, S.} and {Miville-Desch\^enes, M.-A.} and {Moneti, A.} and {Montier, L.} and {Morgante, G.} and {Mortlock, D.} and {Moss, A.} and {Munshi, D.} and {Murphy, J. A.} and {Naselsky, P.} and {Nati, F.} and {Natoli, P.} and {Netterfield, C. B.} and {N\o{}rgaard-Nielsen, H. U.} and {Noviello, F.} and {Novikov, D.} and {Novikov, I.} and {Oxborrow, C. A.} and {Paci, F.} and {Pagano, L.} and {Pajot, F.} and {Paladini, R.} and {Paoletti, D.} and {Partridge, B.} and {Pasian, F.} and {Patanchon, G.} and {Pearson, T. J.} and {Perdereau, O.} and {Perotto, L.} and {Perrotta, F.} and {Pettorino, V.} and {Piacentini, F.} and {Piat, M.} and {Pierpaoli, E.} and {Pietrobon, D.} and {Plaszczynski, S.} and {Pointecouteau, E.} and {Polenta, G.} and {Popa, L.} and {Pratt, G. W.} and {Pr\'ezeau, G.} and {Prunet, S.} and {Puget, J.-L.} and {Rachen, J. P.} and {Reach, W. T.} and {Rebolo, R.} and {Reinecke, M.} and {Remazeilles, M.} and {Renault, C.} and {Renzi, A.} and {Ristorcelli, I.} and {Rocha, G.} and {Rosset, C.} and {Rossetti, M.} and {Roudier, G.} and {Rouill\'e d\'{}Orfeuil, B.} and {Rowan-Robinson, M.} and {Rubi\~no-Mart\'{\i}n, J. A.} and {Rusholme, B.} and {Said, N.} and {Salvatelli, V.} and {Salvati, L.} and {Sandri, M.} and {Santos, D.} and {Savelainen, M.} and {Savini, G.} and {Scott, D.} and {Seiffert, M. D.} and {Serra, P.} and {Shellard, E. P. S.} and {Spencer, L. D.} and {Spinelli, M.} and {Stolyarov, V.} and {Stompor, R.} and {Sudiwala, R.} and {Sunyaev, R.} and {Sutton, D.} and {Suur-Uski, A.-S.} and {Sygnet, J.-F.} and {Tauber, J. A.} and {Terenzi, L.} and {Toffolatti, L.} and {Tomasi, M.} and {Tristram, M.} and {Trombetti, T.} and {Tucci, M.} and {Tuovinen, J.} and {T\"urler, M.} and {Umana, G.} and {Valenziano, L.} and {Valiviita, J.} and {Van Tent, F.} and {Vielva, P.} and {Villa, F.} and {Wade, L. A.} and {Wandelt, B. D.} and {Wehus, I. K.} and {White, M.} and {White, S. D. M.} and {Wilkinson, A.} and {Yvon, D.} and {Zacchei, A.} and {Zonca, A.}},
	title = {Planck 2015 results - XIII. Cosmological parameters},
	DOI= "10.1051/0004-6361/201525830",
	url= "https://doi.org/10.1051/0004-6361/201525830",
	journal = {A\&A},
	year = 2016,
	volume = 594,
	pages = "A13",
}

@article{knollmann&knebe2009,
doi = {10.1088/0067-0049/182/2/608},
url = {https://dx.doi.org/10.1088/0067-0049/182/2/608},
year = {2009},
month = {may},
publisher = {The American Astronomical Society},
volume = {182},
number = {2},
pages = {608},
author = {Steffen R. Knollmann and Alexander Knebe},
title = {Ahf: AMIGA'S HALO FINDER},
journal = {The Astrophysical Journal Supplement Series}
}

@article{yang2018,
doi = {10.3847/1538-4357/aac2ce},
url = {https://dx.doi.org/10.3847/1538-4357/aac2ce},
year = {2018},
month = {jun},
publisher = {The American Astronomical Society},
volume = {860},
number = {1},
pages = {30},
author = {Xiaohu Yang and Youcai Zhang and Huiyuan Wang and Chengze Liu and Tianhuan Lu and Shijie Li and Feng Shi and Y. P. Jing and H. J. Mo and Frank C. van den Bosch and Xi Kang and Weiguang Cui and Hong Guo and Guoliang Li and S. H. Lim and Yi Lu and Wentao Luo and Chengliang Wei and Lei Yang},
title = {ELUCID. V. Lighting Dark Matter Halos with Galaxies},
journal = {The Astrophysical Journal}
}

@article{Loken_2010,
doi = {10.1088/1742-6596/256/1/012026},
url = {https://dx.doi.org/10.1088/1742-6596/256/1/012026},
year = {2010},
month = {nov},
publisher = {},
volume = {256},
number = {1},
pages = {012026},
author = {Chris Loken and Daniel Gruner and Leslie Groer and Richard Peltier and Neil Bunn and Michael Craig and Teresa Henriques and Jillian Dempsey and Ching-Hsing Yu and Joseph Chen and L Jonathan Dursi and Jason Chong and Scott Northrup and Jaime Pinto and Neil Knecht and Ramses Van Zon},
title = {SciNet: Lessons Learned from Building a Power-efficient Top-20 System and Data Centre},
journal = {Journal of Physics: Conference Series}
}

@ARTICLE{Hough2023,
       author = {{Hough}, Renier T. and {Rennehan}, Douglas and {Kobayashi}, Chiaki and {Loubser}, S. Ilani and {Dav{\'e}}, Romeel and {Babul}, Arif and {Cui}, Weiguang},
        title = "{SIMBA-C: an updated chemical enrichment model for galactic chemical evolution in the SIMBA simulation}",
      journal = {\mnras},
         year = 2023,
        month = oct,
       volume = {525},
       number = {1},
        pages = {1061-1076},
          doi = {10.1093/mnras/stad2394},
archivePrefix = {arXiv},
       eprint = {2308.03436},
 primaryClass = {astro-ph.GA},
       adsurl = {https://ui.adsabs.harvard.edu/abs/2023MNRAS.525.1061H}
}

@ARTICLE{Dave2019,
       author = {{Dav{\'e}}, Romeel and {Angl{\'e}s-Alc{\'a}zar}, Daniel and {Narayanan}, Desika and {Li}, Qi and {Rafieferantsoa}, Mika H. and {Appleby}, Sarah},
        title = "{SIMBA: Cosmological simulations with black hole growth and feedback}",
      journal = {\mnras},
         year = 2019,
        month = jun,
       volume = {486},
       number = {2},
        pages = {2827-2849},
          doi = {10.1093/mnras/stz937},
archivePrefix = {arXiv},
       eprint = {1901.10203},
 primaryClass = {astro-ph.GA},
       adsurl = {https://ui.adsabs.harvard.edu/abs/2019MNRAS.486.2827D}
}

@ARTICLE{kravtsov2012,
    author = {{Kravtsov}, A.~V. and {Borgani}, S.},
    title = "{Formation of Galaxy Clusters}",
    journal = {\araa},
    year = 2012,
    volume = {50},
    pages = {353-409},
    doi = {10.1146/annurev-astro-081811-125502}
}

@ARTICLE{chiang2013,
    author = {{Chiang}, Y.-K. and {Overzier}, R. and {Gebhardt}, K.},
    title = "{Ancient Light from Young Cosmic Cities: Physical and Observational Signatures of Galaxy Proto-clusters}",
    journal = {\apj},
    year = 2013,
    month = dec,
    volume = {779},
    eid = {127},
    pages = {127},
    doi = {10.1088/0004-637X/779/2/127}
}

@ARTICLE{overzier2016,
    author = {{Overzier}, R.~A.},
    title = "{The Realm of the Galaxy Protoclusters. A Review}",
    journal = {\aapr},
    year = 2016,
    volume = {24},
    eid = {14},
    pages = {14},
    doi = {10.1007/s00159-016-0100-3}
}

@ARTICLE{chiang2015,
    author = {{Chiang}, Y.-K. and {Overzier}, R.~A. and {Gebhardt}, K. and {Finkelstein}, S.~L. and {Chiang}, C.-T. and {Hill}, G.~J. and {Blanc}, G.~A. and {Drory}, N. and {Chonis}, T.~S. and {Zeimann}, G.~R. and {Hagen}, A. and {Schneider}, D.~P. and {Jogee}, S. and {Ciardullo}, R. and {Gronwall}, C.},
    title = "{Surveying Galaxy Proto-clusters in Emission: A Large-scale Structure at $z=2.44$ and the Outlook for HETDEX}",
    journal = {\apj},
    year = 2015,
    month = jul,
    volume = {808},
    eid = {37},
    pages = {37},
    doi = {10.1088/0004-637X/808/1/37}
}

@ARTICLE{muldrew2018,
    author = {{Muldrew}, S.~I. and {Hatch}, N.~A. and {Cooke}, E.~A.},
    title = "{Galaxy Evolution in Protoclusters}",
    journal = {\mnras},
    year = 2018,
    volume = {473},
    pages = {2335-2347},
    doi = {10.1093/mnras/stx2454}
}

@ARTICLE{castignani2014,
    author = {{Castignani}, G. and {Chiaberge}, M. and {Celotti}, A. and {Norman}, C.},
    title = "{A New Method to Search for High-redshift Clusters Using Photometric Redshifts}",
    journal = {\apj},
    year = 2014,
    month = sep,
    volume = {792},
    eid = {113},
    pages = {113},
    doi = {10.1088/0004-637X/792/2/113}
}

@ARTICLE{chiang2014,
    author = {{Chiang}, Y.-K. and {Overzier}, R. and {Gebhardt}, K.},
    title = "{Discovery of a Large Number of Candidate Protoclusters Traced by $\sim 15$ Mpc-scale Galaxy Overdensities in COSMOS}",
    journal = {\apjl},
    year = 2014,
    month = feb,
    volume = {782},
    eid = {L3},
    pages = {L3},
    doi = {10.1088/2041-8205/782/1/L3}
}

@ARTICLE{toshikawa2018,
    author = {{Toshikawa}, J. and {Uchiyama}, H. and {Kashikawa}, N. and {Ouchi}, M. and {Overzier}, R. and {Ono}, Y. and {Harikane}, Y. and {Ishikawa}, S. and {Kodama}, T. and {Matsuda}, Y. and others},
    title = "{GOLDRUSH. III. A systematic search for protoclusters at $z \sim 4$ based on the $>100\,{\rm deg}^{2}$ area}",
    journal = {\pasj},
    year = 2018,
    volume = {70},
    eid = {S12},
    pages = {S12},
    doi = {10.1093/pasj/psx102}
}

@ARTICLE{li2022,
    author = {{Li}, Q. and {Yang}, X. and {Liu}, C. and {Jing}, Y. and {He}, M. and {Huang}, J.-S. and {Dai}, Y.~S. and {Sawicki}, M. and {Arnouts}, S. and {Gwyn}, S. and others},
    title = "{Groups and Protocluster Candidates in the CLAUDS and HSC-SSP Joint Deep Surveys}",
    journal = {\apj},
    year = 2022,
    month = jul,
    volume = {933},
    eid = {9},
    pages = {9},
    doi = {10.3847/1538-4357/ac6f4a}
}

@ARTICLE{brinch2023,
    author = {{Brinch}, M. and {Greve}, T.~R. and {Weaver}, J.~R. and {Brammer}, G. and {Ilbert}, O. and {Shuntov}, M. and {Jin}, S. and {Liu}, D. and {Gim{\'e}nez-Arteaga}, C. and {Casey}, C.~M. and others},
    title = "{COSMOS2020: Identification of High-z Protocluster Candidates in COSMOS}",
    journal = {\apj},
    year = 2023,
    month = feb,
    volume = {943},
    eid = {153},
    pages = {153},
    doi = {10.3847/1538-4357/ac9d96}
}

@ARTICLE{calvi2023,
    author = {{Calvi}, R. and {Castignani}, G. and {Dannerbauer}, H.},
    title = "{Bright submillimeter galaxies do trace galaxy protoclusters?}",
    journal = {\aap},
    year = 2023,
    month = oct,
    volume = {678},
    eid = {A15},
    pages = {A15},
    doi = {10.1051/0004-6361/202346200}
}

@ARTICLE{gully2024,
    author = {{Gully}, H. and {Hatch}, N. and {Bah{\'e}}, Y. and {Balogh}, M. and {Bolzonella}, M. and {Cooper}, M.~C. and {Muzzin}, A. and {Pozzetti}, L. and {Rudnick}, G. and {Vulcani}, B. and {Wilson}, G.},
    title = "{Spitzer-selected $z > 1.3$ protocluster candidates in the LSST Deep Drilling Fields}",
    journal = {\mnras},
    year = 2024,
    volume = {527},
    pages = {10680-10696},
    doi = {10.1093/mnras/stad3885}
}

@ARTICLE{bohringer2025,
    author = {{B{\"o}hringer}, H. and {Chon}, G. and {Cucciati}, O. and {Dannerbauer}, H. and {Bolzonella}, M. and {De Lucia}, G. and {Cappi}, A. and {Moscardini}, L. and {Giocoli}, C. and {Castignani}, G. and others},
    title = "{Euclid preparation. LV. Exploring the properties of proto-clusters in the Simulated Euclid Wide Survey}",
    journal = {\aap},
    year = 2025,
    volume = {693},
    eid = {A59},
    pages = {A59},
    doi = {10.1051/0004-6361/202451683}
}

@ARTICLE{li2025,
    author = {{Li}, Q. and {Conselice}, C.~J. and {Sarron}, F. and {Harvey}, T. and {Austin}, D. and {Adams}, N. and {Trussler}, J.~A.~A. and {Duan}, Q. and {Ferreira}, L. and {Westcott}, L. and others},
    title = "{EPOCHS paper -- X. Environmental effects on Galaxy formation and protocluster Galaxy candidates at $4.5 < z < 10$ from JWST observations}",
    journal = {\mnras},
    year = 2025,
    volume = {539},
    pages = {1796-1819},
    doi = {10.1093/mnras/staf543}
}

@ARTICLE{vicentin2025a,
    author = {{Vicentin}, M.~C. and {Araya-Araya}, P. and {Sodr{\'e}}, L., Jr. and {Strauss}, M.~A.},
    title = "{Selecting Clusters and Protoclusters via Stellar Mass Density. I. Method and Tests on Mock HSC-SSP Catalogs}",
    journal = {\apj},
    year = 2025,
    volume = {992},
    eid = {52},
    pages = {52},
    doi = {10.3847/1538-4357/adf9aa}
}

@ARTICLE{vicentin2025b,
    author = {{Vicentin}, M.~C. and {Sodr{\'e}}, L., Jr. and {Strauss}, M.~A. and {de Lima}, E.~V.~R. and {Araya-Araya}, P.},
    title = "{Selecting Clusters and Protoclusters via Stellar Mass Density. II. Application to HSC-SSP Observations}",
    journal = {\apj},
    year = 2025,
    volume = {992},
    eid = {53},
    pages = {53},
    doi = {10.3847/1538-4357/adf9ab}
}

@ARTICLE{laureijs2011,
    author = {{Laureijs}, R. and {Amiaux}, J. and {Arduini}, S. and {Augu{\`e}res}, J.-L. and {Brinchmann}, J. and {Cole}, R. and {Cropper}, M. and {Dabin}, C. and {Duvet}, L. and {Ealet}, A. and others},
    title = "{Euclid Definition Study Report}",
    journal = {arXiv e-prints},
    year = 2011,
    month = oct,
    eid = {arXiv:1110.3193},
    pages = {arXiv:1110.3193},
    archivePrefix = {arXiv},
    eprint = {1110.3193},
    primaryClass = {astro-ph.CO}
}

@ARTICLE{ivezic2019,
    author = {{Ivezi{\'c}}, {\v Z}. and {Kahn}, S.~M. and {Tyson}, J.~A. and {Abel}, B. and {Acosta}, E. and {Allsman}, R. and {Alonso}, D. and {AlSayyad}, Y. and {Anderson}, S.~F. and {Andrew}, J. and others},
    title = "{LSST: From Science Drivers to Reference Design and Anticipated Data Products}",
    journal = {\apj},
    year = 2019,
    month = mar,
    volume = {873},
    eid = {111},
    pages = {111},
    doi = {10.3847/1538-4357/ab042c}
}

@ARTICLE{aihara2018,
    author = {{Aihara}, H. and {Arimoto}, N. and {Armstrong}, R. and {Arnouts}, S. and {Bahcall}, N.~A. and {Bickerton}, S. and {Bosch}, J. and {Bundy}, K. and {Capak}, P.~L. and {Chan}, J.~H.~H. and others},
    title = "{The Hyper Suprime-Cam SSP Survey: Overview and survey design}",
    journal = {\pasj},
    year = 2018,
    month = jan,
    volume = {70},
    eid = {S4},
    pages = {S4},
    doi = {10.1093/pasj/psx066}
}

@ARTICLE{takada2014,
    author = {{Takada}, M. and {Ellis}, R.~S. and {Chiba}, M. and {Greene}, J.~E. and {Aihara}, H. and {Arimoto}, N. and {Bundy}, K. and {Cohen}, J. and {Dor{\'e}}, O. and {Graves}, G. and others},
    title = "{Extragalactic science, cosmology, and Galactic archaeology with the Subaru Prime Focus Spectrograph}",
    journal = {\pasj},
    year = 2014,
    month = feb,
    volume = {66},
    eid = {R1},
    pages = {R1},
    doi = {10.1093/pasj/pst019}
}

@ARTICLE{gardner2006,
    author = {{Gardner}, J.~P. and {Mather}, J.~C. and {Clampin}, M. and {Doyon}, R. and {Greenhouse}, M.~A. and {Hammel}, H.~B. and {Hutchings}, J.~B. and {Jakobsen}, P. and {Lilly}, S.~J. and {Long}, K.~S. and others},
    title = "{The James Webb Space Telescope}",
    journal = {\ssr},
    year = 2006,
    month = apr,
    volume = {123},
    pages = {485-606},
    doi = {10.1007/s11214-006-8315-7}
}

@ARTICLE{lovell2018,
    author = {{Lovell}, C.~C. and {Thomas}, P.~A. and {Wilkins}, S.~M.},
    title = "{Characterising and identifying galaxy protoclusters}",
    journal = {\mnras},
    year = 2018,
    month = mar,
    volume = {474},
    number = {4},
    pages = {4612-4625},
    doi = {10.1093/mnras/stx3090},
    archivePrefix = {arXiv},
    eprint = {1710.02148},
    primaryClass = {astro-ph.GA}
}

@ARTICLE{zhang2026,
       author = {{Zhang}, Y.~H. and {Dannerbauer}, H. and
                 {P{\'e}rez-Mart{\'i}nez}, J.~M. and {Koyama}, Y. and
                 {Zheng}, X.~Z. and {Calvi}, R. and {Chen}, Z. and
                 {Daikuhara}, K. and {De Breuck}, C. and
                 {D'Eugenio}, C. and {Emonts}, B.~H.~C. and
                 {Jin}, S. and {Kodama}, T. and {Lehnert}, M.~D. and
                 {Nadolny}, J. and {Naufal}, A. and
                 {P{\'e}rez-Gonz{\'a}lez}, P.~G.},
        title = "{JWST reveals extended stellar disks for ALMA-bright dusty star-forming galaxies in the Spiderweb protocluster}",
      journal = {\aap},
         year = 2026,
        month = feb,
       volume = {706},
          eid = {A9},
        pages = {A9},
          doi = {10.1051/0004-6361/202556611},
archivePrefix = {arXiv},
       eprint = {2512.07050},
 primaryClass = {astro-ph.GA},
       adsurl = {https://ui.adsabs.harvard.edu/abs/2026A%26A...706A...9Z}
}

@ARTICLE{muldrew2015,
       author = {{Muldrew}, Stuart I. and
                 {Hatch}, Nina A. and
                 {Cooke}, Elizabeth A.},
        title = "{What are protoclusters? -- Defining high-redshift galaxy clusters and protoclusters}",
      journal = {\mnras},
         year = 2015,
        month = sep,
       volume = {452},
       number = {3},
        pages = {2528--2539},
          doi = {10.1093/mnras/stv1449},
archivePrefix = {arXiv},
       eprint = {1506.08835},
 primaryClass = {astro-ph.CO},
       adsurl = {https://ui.adsabs.harvard.edu/abs/2015MNRAS.452.2528M}
}

@ARTICLE{gomez2025,
    author = {{G{\'o}mez}, Jonathan S. and {Hough}, Tomas and {Jim{\'e}nez Mu{\~n}oz}, Alejandro and {Yepes}, Gustavo and {Cui}, Weiguang and {Cora}, Sof{\'i}a A.},
    title = "{The Three Hundred Project: A fast semi-analytic model emulator of hydrodynamical galaxy cluster simulations}",
    journal = {\aap},
    year = 2025,
    volume = {697},
    eid = {A171},
    pages = {A171},
    doi = {10.1051/0004-6361/202554064},
    archivePrefix = {arXiv},
    eprint = {2504.03519},
    primaryClass = {astro-ph.GA}
}

@ARTICLE{matsuda2005,
    author = {{Matsuda}, Y. and {Yamada}, T. and {Hayashino}, T. and {Tamura}, H. and {Yamauchi}, R. and {Murayama}, T. and {Nagao}, T. and {Ohta}, K. and {Okamura}, S. and {Ouchi}, M. and {Shimasaku}, K. and {Shioya}, Y. and {Taniguchi}, Y.},
    title = "{Large-Scale Filamentary Structure around the Protocluster at Redshift $z=3.1$}",
    journal = {\apjl},
    year = 2005,
    volume = {634},
    pages = {L125-L128},
    doi = {10.1086/499071},
    archivePrefix = {arXiv},
    eprint = {astro-ph/0510762}
}

@ARTICLE{polletta2021,
  author = {{Polletta}, M. and {Soucail}, G. and {Dole}, H. and {Lehnert}, M. D. and {Pointecouteau}, E. and {Vietri}, G. and {Scodeggio}, M. and {Montier}, L. and {Koyama}, Y. and {Lagache}, G. and {Frye}, B. L. and {Cusano}, F. and {Fumana}, M.},
  title = {{Spectroscopic observations of PHz G237.01+42.50: A galaxy protocluster at z = 2.16 in the Cosmos field}},
  journal = {\aap},
  year = {2021},
  volume = {654},
  pages = {A121},
  doi = {10.1051/0004-6361/202140612},
  url = {https://doi.org/10.1051/0004-6361/202140612}
}

@ARTICLE{shi2021,
  author = {{Shi}, Dong Dong and {Cai}, Zheng and {Fan}, Xiaohui and {Zheng}, Xian Zhong and {Huang}, Yun-Hsin and {Xu}, Jiachuan},
  title = {{Spectroscopic Confirmation of Two Extremely Massive Protoclusters, BOSS1244 and BOSS1542, at z = 2.24}},
  journal = {\apj},
  year = {2021},
  volume = {915},
  pages = {32},
  doi = {10.3847/1538-4357/abfec0},
  url = {https://doi.org/10.3847/1538-4357/abfec0}
}

@ARTICLE{lacaille2019,
  author = {{Lacaille}, Kevin M and {Chapman}, Scott C and {Smail}, Ian and {Steidel}, C C and {Blain}, A W and {Geach}, J and {Golob}, A and {Gurwell}, M and {Ivison}, R J and {Reddy}, N and {Sawicki}, M},
  title = {{Two sub-millimetre bright protoclusters bounding the epoch of peak star-formation activity}},
  journal = {\mnras},
  year = {2019},
  volume = {488},
  pages = {1790-1812},
  doi = {10.1093/mnras/stz1742},
  url = {https://doi.org/10.1093/mnras/stz1742}
}

@ARTICLE{cai2017,
  author = {{Cai}, Zheng and {Fan}, Xiaohui and {Bian}, Fuyan and {Zabludoff}, Ann and {Yang}, Yujin and {Prochaska}, J. Xavier and {McGreer}, Ian and {Zheng}, Zhen-Ya and {Kashikawa}, Nobunari and {Wang}, Ran and {Frye}, Brenda and {Green}, Richard and {Jiang}, Linhua},
  title = {{Mapping the Most Massive Overdensities through Hydrogen (MAMMOTH). II. Discovery of the Extremely Massive Overdensity BOSS1441 at z = 2.32}},
  journal = {\apj},
  year = {2017},
  volume = {839},
  pages = {131},
  doi = {10.3847/1538-4357/aa6a1a},
  url = {https://doi.org/10.3847/1538-4357/aa6a1a}
}

@ARTICLE{lee2017,
  author = {{Lee}, Minju M. and {Tanaka}, Ichi and {Kawabe}, Ryohei and {Kohno}, Kotaro and {Kodama}, Tadayuki and {Kajisawa}, Masaru and {Yun}, Min S. and {Nakanishi}, Kouichiro and {Iono}, Daisuke and {Tamura}, Yoichi and {Hatsukade}, Bunyo and {Umehata}, Hideki and {Saito}, Toshiki and {Izumi}, Takuma and {Aretxaga}, Itziar and {Tadaki}, Ken-ichi and {Zeballos}, Milagros and {Ikarashi}, Soh and {Wilson}, Grant W. and {Hughes}, David H. and {Ivison}, R. J.},
  title = {{A Radio-to-mm Census of Star-forming Galaxies in Protocluster 4C23.56 at Z = 2.5: Gas Mass and Its Fraction Revealed with ALMA}},
  journal = {\apj},
  year = {2017},
  volume = {842},
  pages = {55},
  doi = {10.3847/1538-4357/aa74c2},
  url = {https://doi.org/10.3847/1538-4357/aa74c2}
}

@ARTICLE{venemans2007,
  author = {{Venemans}, B. P. and {R{\"{o}}ttgering}, H. J. A. and {Miley}, G. K. and {van Breugel}, W. J. M. and {De Breuck}, C. and {Kurk}, J. D. and {Pentericci}, L. and {Stanford}, S. A. and {Overzier}, R. A. and {Croft}, S. and {Ford}, H.},
  title = {{Protoclusters associated with z > 2 radio galaxies}},
  journal = {\aap},
  year = {2007},
  volume = {461},
  pages = {823-845},
  doi = {10.1051/0004-6361:20053941},
  url = {https://doi.org/10.1051/0004-6361:20053941}
}

@ARTICLE{cucciati2014,
  author = {{Cucciati}, O. and {Zamorani}, G. and {Lemaux}, B. C. and {Bardelli}, S. and {Cimatti}, A. and {Le F{\`{e}}vre}, O. and {Cassata}, P. and {Garilli}, B. and {Le Brun}, V. and {Maccagni}, D. and {Pentericci}, L. and {Tasca}, L. A. M. and {Thomas}, R. and {Vanzella}, E. and {Zucca}, E. and {Amorin}, R. and {Capak}, P. and {Cassar{\`{a}}}, L. P. and {Castellano}, M. and {Cuby}, J. G. and {de la Torre}, S. and {Durkalec}, A. and {Fontana}, A. and {Giavalisco}, M. and {Grazian}, A. and {Hathi}, N. P. and {Ilbert}, O. and {Moreau}, C. and {Paltani}, S. and {Ribeiro}, B. and {Salvato}, M. and {Schaerer}, D. and {Scodeggio}, M. and {Sommariva}, V. and {Talia}, M. and {Taniguchi}, Y. and {Tresse}, L. and {Vergani}, D. and {Wang}, P. W. and {Charlot}, S. and {Contini}, T. and {Fotopoulou}, S. and {L{\'{o}}pez-Sanjuan}, C. and {Mellier}, Y. and {Scoville}, N.},
  title = {{Discovery of a rich proto-cluster at z = 2.9 and associated diffuse cold gas in the VIMOS Ultra-Deep Survey (VUDS)}},
  journal = {\aap},
  year = {2014},
  volume = {570},
  pages = {A16},
  doi = {10.1051/0004-6361/201423811},
  url = {https://doi.org/10.1051/0004-6361/201423811}
}

@ARTICLE{steidel1998,
  author = {{Steidel}, Charles C. and {Adelberger}, Kurt L. and {Dickinson}, Mark and {Giavalisco}, Mauro and {Pettini}, Max and {Kellogg}, Melinda},
  title = {{A Large Structure of Galaxies at Redshiftz$\sim$ 3 and Its Cosmological Implications}},
  journal = {\apj},
  year = {1998},
  volume = {492},
  pages = {428-438},
  doi = {10.1086/305073},
  url = {https://doi.org/10.1086/305073}
}

@ARTICLE{lemaux2014,
  author = {{Lemaux}, B. C. and {Cucciati}, O. and {Tasca}, L. A. M. and {Le F{\`{e}}vre}, O. and {Zamorani}, G. and {Cassata}, P. and {Garilli}, B. and {Le Brun}, V. and {Maccagni}, D. and {Pentericci}, L. and {Thomas}, R. and {Vanzella}, E. and {Zucca}, E. and {Amor{\'{i}}n}, R. and {Bardelli}, S. and {Capak}, P. and {Cassar{\`{a}}}, L. P. and {Castellano}, M. and {Cimatti}, A. and {Cuby}, J. G. and {de la Torre}, S. and {Durkalec}, A. and {Fontana}, A. and {Giavalisco}, M. and {Grazian}, A. and {Hathi}, N. P. and {Ilbert}, O. and {Moreau}, C. and {Paltani}, S. and {Ribeiro}, B. and {Salvato}, M. and {Schaerer}, D. and {Scodeggio}, M. and {Sommariva}, V. and {Talia}, M. and {Taniguchi}, Y. and {Tresse}, L. and {Vergani}, D. and {Wang}, P. W. and {Charlot}, S. and {Contini}, T. and {Fotopoulou}, S. and {Gal}, R. R. and {Kocevski}, D. D. and {L{\'{o}}pez-Sanjuan}, C. and {Lubin}, L. M. and {Mellier}, Y. and {Sadibekova}, T. and {Scoville}, N.},
  title = {{VIMOS Ultra-Deep Survey (VUDS): Witnessing the assembly of a massive cluster atz~ 3.3}},
  journal = {\aap},
  year = {2014},
  volume = {572},
  pages = {A41},
  doi = {10.1051/0004-6361/201423828},
  url = {https://doi.org/10.1051/0004-6361/201423828}
}

@ARTICLE{mcconachie2022,
  author = {{McConachie}, Ian and {Wilson}, Gillian and {Forrest}, Ben and {Marsan}, Z. Cemile and {Muzzin}, Adam and {Cooper}, M. C. and {Annunziatella}, Marianna and {Marchesini}, Danilo and {Chan}, Jeffrey C. C. and {Gomez}, Percy and {Abdullah}, Mohamed H. and {Saracco}, Paolo and {Nantais}, Julie},
  title = {{Spectroscopic Confirmation of a Protocluster at z = 3.37 with a High Fraction of Quiescent Galaxies}},
  journal = {\apj},
  year = {2022},
  volume = {926},
  pages = {37},
  doi = {10.3847/1538-4357/ac2b9f},
  url = {https://doi.org/10.3847/1538-4357/ac2b9f}
}

@ARTICLE{toshikawa2016,
  author = {{Toshikawa}, Jun and {Kashikawa}, Nobunari and {Overzier}, Roderik and {Malkan}, Matthew A. and {Furusawa}, Hisanori and {Ishikawa}, Shogo and {Onoue}, Masafusa and {Ota}, Kazuaki and {Tanaka}, Masayuki and {Niino}, Yuu and {Uchiyama}, Hisakazu},
  title = {{A SYSTEMATIC SURVEY OF PROTOCLUSTERS AT z $\sim$ 3--6 IN THE CFHTLS DEEP FIELDS}},
  journal = {\apj},
  year = {2016},
  volume = {826},
  pages = {114},
  doi = {10.3847/0004-637x/826/2/114},
  url = {https://doi.org/10.3847/0004-637x/826/2/114}
}

@ARTICLE{toshikawa2025,
  author = {{Toshikawa}, Jun and {Wuyts}, Stijn and {Kashikawa}, Nobunari and {Uchiyama}, Hisakazu and {Bremer}, Malcolm and {Sawicki}, Marcin and {Ono}, Yoshiaki and {Kubo}, Mariko and {Ito}, Kei and {Ali}, Sadman S},
  title = {{Galaxy properties from the outskirts to the core of a protocluster at z = 3.70}},
  journal = {\mnras},
  year = {2025},
  volume = {537},
  pages = {3561-3574},
  doi = {10.1093/mnras/staf197},
  url = {https://doi.org/10.1093/mnras/staf197}
}

@ARTICLE{dey2016,
  author = {{Dey}, Arjun and {Lee}, Kyoung-Soo and {Reddy}, Naveen and {Cooper}, Michael and {Inami}, Hanae and {Hong}, Sungryong and {Gonzalez}, Anthony H. and {Jannuzi}, Buell T.},
  title = {{SPECTROSCOPIC CONFIRMATION OF A PROTOCLUSTER AT z $\simeq$ 3.786}},
  journal = {\apj},
  year = {2016},
  volume = {823},
  pages = {11},
  doi = {10.3847/0004-637x/823/1/11},
  url = {https://doi.org/10.3847/0004-637x/823/1/11}
}

@ARTICLE{oteo2018,
  author = {{Oteo}, I. and {Ivison}, R. J. and {Dunne}, L. and {Manilla-Robles}, A. and {Maddox}, S. and {Lewis}, A. J. R. and {Zotti}, G. de and {Bremer}, M. and {Clements}, D. L. and {Cooray}, A. and {Dannerbauer}, H. and {Eales}, S. and {Greenslade}, J. and {Omont}, A. and {Perez--Fourn{\'{o}}n}, I. and {Riechers}, D. and {Scott}, D. and {van der Werf}, P. and {Weiss}, A. and {Zhang}, Z-Y.},
  title = {{An Extreme Protocluster of Luminous Dusty Starbursts in the Early Universe}},
  journal = {\apj},
  year = {2018},
  volume = {856},
  pages = {72},
  doi = {10.3847/1538-4357/aaa1f1},
  url = {https://doi.org/10.3847/1538-4357/aaa1f1}
}

@ARTICLE{miller2018,
  author = {{Miller}, T. B. and {Chapman}, S. C. and {Aravena}, M. and {Ashby}, M. L. N. and {Hayward}, C. C. and {Vieira}, J. D. and {Wei}, A. and {Babul}, A. and {B{\'{e}}thermin}, M. and {Bradford}, C. M. and {Brodwin}, M. and {Carlstrom}, J. E. and {Chen}, Chian-Chou and {Cunningham}, D. J. M. and {De Breuck}, C. and {Gonzalez}, A. H. and {Greve}, T. R. and {Harnett}, J. and {Hezaveh}, Y. and {Lacaille}, K. and {Litke}, K. C. and {Ma}, J. and {Malkan}, M. and {Marrone}, D. P. and {Morningstar}, W. and {Murphy}, E. J. and {Narayanan}, D. and {Pass}, E. and {Perry}, R. and {Phadke}, K. A. and {Rennehan}, D. and {Rotermund}, K. M. and {Simpson}, J. and {Spilker}, J. S. and {Sreevani}, J. and {Stark}, A. A. and {Strandet}, M. L. and {Strom}, A. L.},
  title = {{A massive core for a cluster of galaxies at a redshift of 4.3}},
  journal = {\nat},
  year = {2018},
  volume = {556},
  pages = {469-472},
  doi = {10.1038/s41586-018-0025-2},
  url = {https://doi.org/10.1038/s41586-018-0025-2}
}

@ARTICLE{helton2024,
  author = {{Helton}, Jakob M. and {Sun}, Fengwu and {Woodrum}, Charity and {Hainline}, Kevin N. and {Willmer}, Christopher N. A. and {Rieke}, Marcia J. and {Rieke}, George H. and {Alberts}, Stacey and {Eisenstein}, Daniel J. and {Tacchella}, Sandro and {Robertson}, Brant and {Johnson}, Benjamin D. and {Baker}, William M. and {Bhatawdekar}, Rachana and {Bunker}, Andrew J. and {Chen}, Zuyi and {Egami}, Eiichi and {Ji}, Zhiyuan and {Maiolino}, Roberto and {Willott}, Chris and {Witstok}, Joris},
  title = {{Identification of High-redshift Galaxy Overdensities in GOODS-N and GOODS-S}},
  journal = {\apj},
  year = {2024},
  volume = {974},
  pages = {41},
  doi = {10.3847/1538-4357/ad6867},
  url = {https://doi.org/10.3847/1538-4357/ad6867}
}

@ARTICLE{harikane2019,
  author = {{Harikane}, Yuichi and {Ouchi}, Masami and {Ono}, Yoshiaki and {Fujimoto}, Seiji and {Donevski}, Darko and {Shibuya}, Takatoshi and {Faisst}, Andreas L. and {Goto}, Tomotsugu and {Hatsukade}, Bunyo and {Kashikawa}, Nobunari and {Kohno}, Kotaro and {Hashimoto}, Takuya and {Higuchi}, Ryo and {Inoue}, Akio K. and {Lin}, Yen-Ting and {Martin}, Crystal L. and {Overzier}, Roderik and {Smail}, Ian and {Toshikawa}, Jun and {Umehata}, Hideki and {Ao}, Yiping and {Chapman}, Scott and {Clements}, David L. and {Im}, Myungshin and {Jing}, Yipeng and {Kawaguchi}, Toshihiro and {Lee}, Chien-Hsiu and {Lee}, Minju M. and {Lin}, Lihwai and {Matsuoka}, Yoshiki and {Marinello}, Murilo and {Nagao}, Tohru and {Onodera}, Masato and {Toft}, Sune and {Wang}, Wei-Hao},
  title = {{SILVERRUSH. VIII. Spectroscopic Identifications of Early Large-scale Structures with Protoclusters over 200 Mpc at z $\sim$ 6--7: Strong Associations of Dusty Star-forming Galaxies}},
  journal = {\apj},
  year = {2019},
  volume = {883},
  pages = {142},
  doi = {10.3847/1538-4357/ab2cd5},
  url = {https://doi.org/10.3847/1538-4357/ab2cd5}
}

@ARTICLE{morishita2025,
  author = {{Morishita}, Takahiro and {Liu}, Zhaoran and {Stiavelli}, Massimo and {Treu}, Tommaso and {Trenti}, Michele and {Chartab}, Nima and {Roberts-Borsani}, Guido and {Vulcani}, Benedetta and {Bergamini}, Pietro and {Castellano}, Marco and {Grillo}, Claudio},
  title = {{Accelerated Emergence of Evolved Galaxies in Early Overdensities at z $\sim$ 5.7}},
  journal = {\apj},
  year = {2025},
  volume = {982},
  pages = {153},
  doi = {10.3847/1538-4357/adb30f},
  url = {https://doi.org/10.3847/1538-4357/adb30f}
}

@ARTICLE{brinch2024,
  author = {{Brinch}, Malte and {Greve}, Thomas R and {Sanders}, David B and {McPartland}, Conor J R and {Chartab}, Nima and {Gillman}, Steven and {Vijayan}, Aswin P and {Lee}, Minju M and {Brammer}, Gabriel and {Casey}, Caitlin M and {Ilbert}, Olivier and {Jin}, Shuowen and {Magdis}, Georgios and {McCracken}, H J and {Sillassen}, Nikolaj B and {Toft}, Sune and {Zavala}, Jorge A},
  title = {{DEIMOS spectroscopy of z = 6 protocluster candidate in COSMOS -- a massive protocluster embedded in a large-scale structure?}},
  journal = {\mnras},
  year = {2024},
  volume = {527},
  pages = {6591-6615},
  doi = {10.1093/mnras/stad3409},
  url = {https://doi.org/10.1093/mnras/stad3409}
}

@ARTICLE{toshikawa2014,
  author = {{Toshikawa}, Jun and {Kashikawa}, Nobunari and {Overzier}, Roderik and {Shibuya}, Takatoshi and {Ishikawa}, Shogo and {Ota}, Kazuaki and {Shimasaku}, Kazuhiro and {Tanaka}, Masayuki and {Hayashi}, Masao and {Niino}, Yuu and {Onoue}, Masafusa},
  title = {{A FIRST SITE OF GALAXY CLUSTER FORMATION: COMPLETE SPECTROSCOPY OF A PROTOCLUSTER AT z = 6.01}},
  journal = {\apj},
  year = {2014},
  volume = {792},
  pages = {15},
  doi = {10.1088/0004-637x/792/1/15},
  url = {https://doi.org/10.1088/0004-637x/792/1/15}
}

@ARTICLE{higuchi2019,
  author = {{Higuchi}, Ryo and {Ouchi}, Masami and {Ono}, Yoshiaki and {Shibuya}, Takatoshi and {Toshikawa}, Jun and {Harikane}, Yuichi and {Kojima}, Takashi and {Chiang}, Yi-Kuan and {Egami}, Eiichi and {Kashikawa}, Nobunari and {Overzier}, Roderik and {Konno}, Akira and {Inoue}, Akio K. and {Hasegawa}, Kenji and {Fujimoto}, Seiji and {Goto}, Tomotsugu and {Ishikawa}, Shogo and {Ito}, Kei and {Komiyama}, Yutaka and {Tanaka}, Masayuki},
  title = {{SILVERRUSH. VII. Subaru/HSC Identifications of Protocluster Candidates at z $\sim$ 6--7: Implications for Cosmic Reionization}},
  journal = {\apj},
  year = {2019},
  volume = {879},
  pages = {28},
  doi = {10.3847/1538-4357/ab2192},
  url = {https://doi.org/10.3847/1538-4357/ab2192}
}

@ARTICLE{hu2021,
  author = {{Hu}, Weida and {Wang}, Junxian and {Infante}, Leopoldo and {Rhoads}, James E. and {Zheng}, Zhen-Ya and {Yang}, Huan and {Malhotra}, Sangeeta and {Barrientos}, L. Felipe and {Jiang}, Chunyan and {Gonz{\'{a}}lez-L{\'{o}}pez}, Jorge and {Prieto}, Gonzalo and {Perez}, Lucia A. and {Hibon}, Pascale and {Galaz}, Gaspar and {Coughlin}, Alicia and {Harish}, Santosh and {Kong}, Xu and {Kang}, Wenyong and {Khostovan}, Ali Ahmad and {Pharo}, John and {Valdes}, Francisco and {Wold}, Isak and {Walker}, Alistair R. and {Zheng}, XianZhong},
  title = {{A Lyman-$\alpha$ protocluster at redshift 6.9}},
  journal = {Nature Astronomy},
  year = {2021},
  volume = {5},
  pages = {485-490},
  doi = {10.1038/s41550-020-01291-y},
  url = {https://doi.org/10.1038/s41550-020-01291-y}
}

@ARTICLE{champagne2025,
  author = {{Champagne}, Jaclyn B. and {Wang}, Feige and {Zhang }, Haowen  and {Yang}, Jinyi and {Fan}, Xiaohui and {Hennawi}, Joseph F. and {Sun}, Fengwu and {Ba{\~{n}}ados}, Eduardo and {Bosman}, Sarah E. I. and {Costa}, Tiago and {Eilers}, Anna-Christina and {Endsley}, Ryan and {Jin}, Xiangyu and {Jun}, Hyunsung D. and {Li}, Mingyu and {Lin}, Xiaojing and {Liu}, Weizhe and {Loiacono}, Federica and {Lupi}, Alessandro and {Mazzucchelli}, Chiara and {Pudoka}, Maria and {Protu{\v{s}}ov{\`{a}}}, Klaudia and {Rojas-Ruiz}, Sof{\'{i}}a and {Tee}, Wei Leong and {Trebitsch}, Maxime and {Venemans}, Bram P. and {Zhuang }, Ming-Yang  and {Zou}, Siwei},
  title = {{A Quasar-anchored Protocluster at z = 6.6 in the ASPIRE Survey. I. Properties of [O iii] Emitters in a 10 Mpc Overdensity Structure}},
  journal = {\apj},
  year = {2025},
  volume = {981},
  pages = {113},
  doi = {10.3847/1538-4357/adb1bd},
  url = {https://doi.org/10.3847/1538-4357/adb1bd}
}

@ARTICLE{bogdan2026,
  author = {{Bogd{\'{a}}n}, {\'{A}}kos and {Schellenberger}, Gerrit and {Li}, Qiong and {Conselice}, Christopher J.},
  title = {{An X-ray-emitting protocluster at z $\simeq$ 5.7 reveals rapid structure growth}},
  journal = {\nat},
  year = {2026},
  volume = {649},
  pages = {1134-1138},
  doi = {10.1038/s41586-025-09973-1},
  url = {https://doi.org/10.1038/s41586-025-09973-1}
}

@ARTICLE{laporte2022,
  author = {{Laporte}, N. and {Zitrin}, A. and {Dole}, H. and {Roberts-Borsani}, G. and {Furtak}, L. J. and {Witten}, C.},
  title = {{A lensed protocluster candidate at z = 7.66 identified in JWST observations of the galaxy cluster SMACS0723-7327}},
  journal = {\aap},
  year = {2022},
  volume = {667},
  pages = {L3},
  doi = {10.1051/0004-6361/202244719},
  url = {https://doi.org/10.1051/0004-6361/202244719}
}

@ARTICLE{hatch2016,
  author = {{Hatch}, N. A. and {Muldrew}, S. I. and {Cooke}, E. A. and {Hartley}, W. G. and {Almaini}, O. and {Simpson}, C. J. and {Conselice}, C. J.},
  title = {{The structure and evolution of a forming galaxy cluster atz= 1.62}},
  journal = {\mnras},
  year = {2016},
  volume = {459},
  pages = {387-401},
  doi = {10.1093/mnras/stw602},
  url = {https://doi.org/10.1093/mnras/stw602}
}

@ARTICLE{santos2014,
  author = {{Santos}, J. S. and {Altieri}, B. and {Tanaka}, M. and {Valtchanov}, I. and {Saintonge}, A. and {Dickinson}, M. and {Foucaud}, S. and {Kodama}, T. and {Rawle}, T. D. and {Tadaki}, K.},
  title = {{Star formation in the cluster CLG0218.3-0510 at z = 1.62 and its large-scale environment: the infrared perspective}},
  journal = {\mnras},
  year = {2014},
  volume = {438},
  pages = {2565-2577},
  doi = {10.1093/mnras/stt2376},
  url = {https://doi.org/10.1093/mnras/stt2376}
}

@ARTICLE{casasola2018,
  author = {{Casasola}, V. and {Magrini}, L. and {Combes}, F. and {Sani}, E. and {Fritz}, J. and {Rodighiero}, G. and {Poggianti}, B. and {Bianchi}, S. and {Liuzzo}, E.},
  title = {{Spectroscopic characterization of the protocluster of galaxies around 7C 1756+6520 at z ~ 1.4}},
  journal = {\aap},
  year = {2018},
  volume = {618},
  pages = {A128},
  doi = {10.1051/0004-6361/201833052},
  url = {https://doi.org/10.1051/0004-6361/201833052}
}

@ARTICLE{cucciati2018_hyperion,
  author = {{Cucciati}, O. and {Lemaux}, B. C. and {Zamorani}, G. and {Le F{\`{e}}vre}, O. and {Tasca}, L. A. M. and {Hathi}, N. P. and {Lee}, K.-G. and {Bardelli}, S. and {Cassata}, P. and {Garilli}, B. and {Le Brun}, V. and {Maccagni}, D. and {Pentericci}, L. and {Thomas}, R. and {Vanzella}, E. and {Zucca}, E. and {Lubin}, L. M. and {Amorin}, R. and {Cassar{\`{a}}}, L. P. and {Cimatti}, A. and {Talia}, M. and {Vergani}, D. and {Koekemoer}, A. and {Pforr}, J. and {Salvato}, M.},
  title = {{The progeny of a cosmic titan: a massive multi-component proto-supercluster in formation at z = 2.45 in VUDS}},
  journal = {\aap},
  year = {2018},
  volume = {619},
  pages = {A49},
  doi = {10.1051/0004-6361/201833655},
  url = {https://doi.org/10.1051/0004-6361/201833655}
}

@ARTICLE{old2020_gogreen,
  author = {{Old}, Lyndsay J and {Balogh}, Michael L and {van der Burg}, Remco F J and {Biviano}, Andrea and {Yee}, Howard K C and {Pintos-Castro}, Irene and {Webb}, Kristi and {Muzzin}, Adam and {Rudnick}, Gregory and {Vulcani}, Benedetta and {Poggianti}, Bianca and {Cooper}, Michael and {Zaritsky}, Dennis and {Cerulo}, Pierluigi and {Wilson}, Gillian and {Chan}, Jeffrey C C and {Lidman}, Chris and {McGee}, Sean and {Demarco}, Ricardo and {Forrest}, Ben and {De Lucia}, Gabriella and {Gilbank}, David and {Kukstas}, Egidijus and {McCarthy}, Ian G and {Jablonka}, Pascale and {Nantais}, Julie and {Noble}, Allison and {Reeves}, Andrew M M and {Shipley}, Heath},
  title = {{The GOGREEN survey: the environmental dependence of the star-forming galaxy main sequence at 1.0 \&lt; z \&lt; 1.5}},
  journal = {\mnras},
  year = {2020},
  volume = {493},
  pages = {5987-6000},
  doi = {10.1093/mnras/staa579},
  url = {https://doi.org/10.1093/mnras/staa579}
}

@ARTICLE{sifon2025_chances,
  author = {{Sif{\'{o}}n}, Crist{\'{o}}bal and {Finoguenov}, Alexis and {Haines}, Christopher P. and {Jaff{\'{e}}}, Yara and {Amrutha}, B. M. and {Demarco}, Ricardo and {Lima}, E. V. R. and {Lima-Dias}, Ciria and {M{\'{e}}ndez-Hern{\'{a}}ndez}, Hugo and {Merluzzi}, Paola and {Monachesi}, Antonela and {Teixeira}, Gabriel S. M. and {Tejos}, Nicolas and {Almeida-Fernandes}, F. and {Araya-Araya}, Pablo and {Argudo-Fern{\'{a}}ndez}, Maria and {Baier-Soto}, Ra{\'{u}}l and {Bilton}, Lawrence E. and {Bom}, C. R. and {Calder{\'{o}}n}, Juan Pablo and {Cassar{\`{a}}}, Letizia P. and {Comparat}, Johan and {Courtois}, H. M. and {DAgo}, Giuseppe and {Dupuy}, Alexandra and {Fritz}, Alexander and {Haack}, Rodrigo F. and {Herpich}, Fabio R. and {Ibar}, E. and {Kuchner}, Ulrike and {Lacerna}, Ivan and {Lopes}, Amanda R. and {Lopez}, Sebastian and {L{\"{o}}sch}, Elismar and {McGee}, Sean and {Mendes de Oliveira}, C. and {Morelli}, Lorenzo and {Moretti}, Alessia and {Pallero}, Diego and {Piraino-Cerda}, Franco and {Pompei}, Emanuela and {Rescigno}, U. and {Smith Castelli}, Anal{\'{i}}a V. and {Smith}, Rory and {Sodr{\'{e}} Jr}, Laerte and {Tempel}, Elmo},
  title = {{CHANCES, the Chilean Cluster Galaxy Evolution Survey: Selection and initial characterisation of clusters and superclusters}},
  journal = {\aap},
  year = {2025},
  volume = {697},
  pages = {A92},
  doi = {10.1051/0004-6361/202452710},
  url = {https://doi.org/10.1051/0004-6361/202452710}
}

@ARTICLE{baxter2025,
  author = {{Baxter}, Devontae C. and {Coil}, Alison L. and {Nadler}, Ethan O. and {Nelson}, Dylan and {Pillepich}, Annalisa and {Forrest}, Ben and {Giddings}, Finn and {Golden-Marx}, Emmet and {Lemaux}, Brian C. and {Sikorski}, Derek},
  title = {{Quantifying the Impact of Incompleteness on Identifying and Interpreting Galaxy Protocluster Populations with the TNG-Cluster Simulation}},
  journal = {\apj},
  year = {2025},
  volume = {990},
  pages = {225},
  doi = {10.3847/1538-4357/adf644},
  url = {https://doi.org/10.3847/1538-4357/adf644}
}

@ARTICLE{morishita2023_glass,
  author = {{Morishita}, Takahiro and {Roberts-Borsani}, Guido and {Treu}, Tommaso and {Brammer}, Gabriel and {Mason}, Charlotte A. and {Trenti}, Michele and {Vulcani}, Benedetta and {Wang}, Xin and {Acebron}, Ana and {Bah{\'{e}}}, Yannick and {Bergamini}, Pietro and {Boyett}, Kristan and {Bradac}, Marusa and {Calabr{\`{o}}}, Antonello and {Castellano}, Marco and {Chen}, Wenlei and {De Lucia}, Gabriella and {Filippenko}, Alexei V. and {Fontana}, Adriano and {Glazebrook}, Karl and {Grillo}, Claudio and {Henry}, Alaina and {Jones}, Tucker and {Kelly}, Patrick L. and {Koekemoer}, Anton M. and {Leethochawalit}, Nicha and {Lu}, Ting-Yi and {Marchesini}, Danilo and {Mascia}, Sara and {Mercurio}, Amata and {Merlin}, Emiliano and {Metha}, Benjamin and {Nanayakkara}, Themiya and {Nonino}, Mario and {Paris}, Diego and {Pentericci}, Laura and {Rosati}, Piero and {Santini}, Paola and {Strait}, Victoria and {Vanzella}, Eros and {Windhorst}, Rogier A. and {Xie}, Lizhi},
  title = {{Early Results from GLASS-JWST. XIV. A Spectroscopically Confirmed Protocluster 650 Million Years after the Big Bang}},
  journal = {\apjl},
  year = {2023},
  volume = {947},
  pages = {L24},
  doi = {10.3847/2041-8213/acb99e},
  url = {https://doi.org/10.3847/2041-8213/acb99e}
}

@ARTICLE{pentericci1998,
  author = {{Pentericci}, L. and {Rottgering}, H. J. A. and {Miley}, G. K. and {Spinrad}, H. and {McCarthy}, P. J. and {van Breugel}, W. J. M. and {Macchetto}, F.},
  title = {{HSTImages of the Extremely Clumpy Radio Galaxy 1138-262 atz= 2.2}},
  journal = {\apj},
  year = {1998},
  volume = {504},
  pages = {139-146},
  doi = {10.1086/306087},
  url = {https://doi.org/10.1086/306087}
}

@ARTICLE{perezmartinez2024,
  author = {{P{\'{e}}rez-Mart{\'{i}}nez}, J M and {Kodama}, T and {Koyama}, Y and {Shimakawa}, R and {Suzuki}, T L and {Daikuhara}, K and {Adachi}, K and {Onodera}, M and {Tanaka}, I},
  title = {{Enhanced star formation and metallicity deficit in the USS 1558-003 forming protocluster at z = 2.53}},
  journal = {\mnras},
  year = {2024},
  volume = {527},
  pages = {10221-10238},
  doi = {10.1093/mnras/stad3805},
  url = {https://doi.org/10.1093/mnras/stad3805}
}

@ARTICLE{shah2024,
  author = {{Shah}, Ekta A and {Lemaux}, Brian and {Forrest}, Benjamin and {Cucciati}, Olga and {Hung}, Denise and {Staab}, Priti and {Hathi}, Nimish and {Lubin}, Lori and {Gal}, Roy R and {Shen}, Lu and {Zamorani}, Giovanni and {Giddings}, Finn and {Bardelli}, Sandro and {Cassara}, Letizia Pasqua and {Cassata}, Paolo and {Contini}, Thierry and {Golden-Marx}, Emmet and {Guaita}, Lucia and {Gururajan}, Gayathri and {Koekemoer}, Anton M and {McLeod}, Derek and {Tasca}, Lidia A M and {Tresse}, Laurence and {Vergani}, Daniela and {Zucca}, Elena},
  title = {{Identification and characterization of six spectroscopically confirmed massive protostructures at 2.5 \&lt; z \&lt; 4.5}},
  journal = {\mnras},
  year = {2024},
  volume = {529},
  pages = {873-882},
  doi = {10.1093/mnras/stae519},
  url = {https://doi.org/10.1093/mnras/stae519}
}

@ARTICLE{overzier2008,
  author = {{Overzier}, Roderik A. and {Bouwens}, R. J. and {Cross}, N. J. G. and {Venemans}, B. P. and {Miley}, G. K. and {Zirm}, A. W. and {Benitez}, N. and {Blakeslee}, J. P. and {Coe}, D. and {Demarco}, R. and {Ford}, H. C. and {Homeier}, N. L. and {Illingworth}, G. D. and {Kurk}, J. D. and {Martel}, A. R. and {Mei}, S. and {Oliveira}, I. and {Rottgering}, H. J. A. and {Tsvetanov}, Z. I. and {Zheng}, W.},
  title = {{Lyman Break Galaxies, Ly$\alpha$ Emitters, and a Radio Galaxy in a Protocluster atz= 4.1}},
  journal = {\apj},
  year = {2008},
  volume = {673},
  pages = {143-162},
  doi = {10.1086/524342},
  url = {https://doi.org/10.1086/524342}
}

@ARTICLE{hayashi2011,
  author = {{Hayashi}, Masao and {Kodama}, Tadayuki and {Koyama}, Yusei and {Tadaki}, Ken-ichi and {Tanaka}, Ichi},
  title = {{Properties of star-forming galaxies in a cluster and its surrounding structure at}},
  journal = {\mnras},
  year = {2011},
  volume = {415},
  pages = {2670-2687},
  doi = {10.1111/j.1365-2966.2011.18892.x},
  url = {https://doi.org/10.1111/j.1365-2966.2011.18892.x}
}

@ARTICLE{papovich2010,
  author = {{Papovich}, C. and {Momcheva}, I. and {Willmer}, C. N. A. and {Finkelstein}, K. D. and {Finkelstein}, S. L. and {Tran}, K.-V. and {Brodwin}, M. and {Dunlop}, J. S. and {Farrah}, D. and {Khan}, S. A. and {Lotz}, J. and {McCarthy}, P. and {McLure}, R. J. and {Rieke}, M. and {Rudnick}, G. and {Sivanandam}, S. and {Pacaud}, F. and {Pierre}, M.},
  title = {{ASPITZER-SELECTED GALAXY CLUSTER ATz= 1.62}},
  journal = {\apj},
  year = {2010},
  volume = {716},
  pages = {1503-1513},
  doi = {10.1088/0004-637x/716/2/1503},
  url = {https://doi.org/10.1088/0004-637x/716/2/1503}
}

@ARTICLE{coogan2018,
  author = {{Coogan}, R T and {Daddi}, E and {Sargent}, M T and {Strazzullo}, V and {Valentino}, F and {Gobat}, R and {Magdis}, G and {Bethermin}, M and {Pannella}, M and {Onodera}, M and {Liu}, D and {Cimatti}, A and {Dannerbauer}, H and {Carollo}, M and {Renzini}, A and {Tremou}, E},
  title = {{Merger driven star-formation activity in Cl J1449+0856 at z=1.99 as seen by ALMA and JVLA}},
  journal = {\mnras},
  year = {2018},
  doi = {10.1093/mnras/sty1446},
  url = {https://doi.org/10.1093/mnras/sty1446}
}

@ARTICLE{smith2019,
  author = {{Smith}, C M A and {Gear}, W K and {Smith}, M W L and {Papageorgiou}, A and {Eales}, S A},
  title = {{Revealing dust-obscured star formation in CLJ1449+0856, a cluster at z = 2}},
  journal = {\mnras},
  year = {2019},
  volume = {486},
  pages = {4304-4319},
  doi = {10.1093/mnras/stz1090},
  url = {https://doi.org/10.1093/mnras/stz1090}
}

@ARTICLE{casey2015,
  author = {{Casey}, C. M. and {Cooray}, A. and {Capak}, P. and {Fu}, H. and {Kovac}, K. and {Lilly}, S. and {Sanders}, D. B. and {Scoville}, N. Z. and {Treister}, E.},
  title = {{A MASSIVE, DISTANT PROTO-CLUSTER AT z = 2.47 CAUGHT IN A PHASE OF RAPID FORMATION?}},
  journal = {\apj},
  year = {2015},
  volume = {808},
  pages = {L33},
  doi = {10.1088/2041-8205/808/2/l33},
  url = {https://doi.org/10.1088/2041-8205/808/2/l33}
}

@ARTICLE{Onions2025,
       author = {{Onions}, Julian and {Pearce}, Frazer and {Knebe}, Alexander and {Gray}, Meghan and {Haggar}, Roan and {Kuchner}, Ulrike and {Contreras-Santos}, Ana and {Yepes}, Gustavo and {Cui}, Weiguang},
        title = "{The life and times of dark matter haloes: what will I be when I grow up?}",
      journal = {\mnras},
         year = 2025,
        month = sep,
       volume = {542},
       number = {2},
        pages = {1477-1485},
          doi = {10.1093/mnras/staf1293},
archivePrefix = {arXiv},
       eprint = {2508.18778},
 primaryClass = {astro-ph.GA},
       adsurl = {https://ui.adsabs.harvard.edu/abs/2025MNRAS.542.1477O}
}

@ARTICLE{Remus2023,
       author = {{Remus}, Rhea-Silvia and {Dolag}, Klaus and {Dannerbauer}, Helmut},
        title = "{The Young and the Wild: What Happens to Protoclusters Forming at Redshift z {\ensuremath{\approx}} 4?}",
      journal = {\apj},
         year = 2023,
        month = jun,
       volume = {950},
       number = {2},
          eid = {191},
        pages = {191},
          doi = {10.3847/1538-4357/accb91},
archivePrefix = {arXiv},
       eprint = {2208.01053},
 primaryClass = {astro-ph.CO},
       adsurl = {https://ui.adsabs.harvard.edu/abs/2023ApJ...950..191R}
}

@ARTICLE{Baxter2026,
       author = {{Baxter}, Devontae C. and {Coil}, Alison L. and {Nadler}, Ethan O and {Shen}, Xuejian and {Vogelsberger}, Mark},
        title = "{What becomes of JWST/NIRCam-selected high-redshift massive galaxies?}",
      journal = {arXiv e-prints},
         year = 2026,
        month = jul,
          eid = {arXiv:2607.08820},
        pages = {arXiv:2607.08820},
          doi = {10.48550/arXiv.2607.08820},
archivePrefix = {arXiv},
       eprint = {2607.08820},
 primaryClass = {astro-ph.GA},
       adsurl = {https://ui.adsabs.harvard.edu/abs/2026arXiv260708820B}
}

@ARTICLE{Witten2026,
       author = {{Witten}, Callum and {Bennett}, Jake S. and {Oesch}, Pascal A. and {Lim}, Seunghwan and {Terp}, Chamilla and {Helton}, Jakob M. and {Heintz}, Kasper E. and {Meyer}, Romain A. and {McClymont}, William and {Herard-Demanche}, Thomas and {Giovinazzo}, Emma},
        title = "{Too many protoclusters? Reconciling the overabundance of cluster progenitors within the first billion years of the Universe}",
      journal = {arXiv e-prints},
         year = 2026,
        month = may,
          eid = {arXiv:2605.28930},
        pages = {arXiv:2605.28930},
          doi = {10.48550/arXiv.2605.28930},
archivePrefix = {arXiv},
       eprint = {2605.28930},
 primaryClass = {astro-ph.GA},
       adsurl = {https://ui.adsabs.harvard.edu/abs/2026arXiv260528930W}
}

@ARTICLE{Cui2020,
       author = {{Cui}, Weiguang and {Qiao}, Jiaqi and {Dav{\'e}}, Romeel and {Knebe}, Alexander and {Peacock}, John A. and {Yepes}, Gustavo},
        title = "{Protoclusters at z = 5.7: a view from the MultiDark galaxies}",
      journal = {Monthly Notices of the Royal Astronomical Society},
         year = 2020,
        month = oct,
       volume = {497},
       number = {4},
        pages = {5220-5228},
          doi = {10.1093/mnras/staa2319},
archivePrefix = {arXiv},
       eprint = {2008.01816},
 primaryClass = {astro-ph.GA},
       adsurl = {https://ui.adsabs.harvard.edu/abs/2020MNRAS.497.5220C}
}

\begin{appendix}
\nolinenumbers

\section{One-dimensional Step~1 parameter slices}
\label{App:Step1Slices}
\begin{figure*}[tbp]
    \centering
    \includegraphics[width=0.46\textwidth]{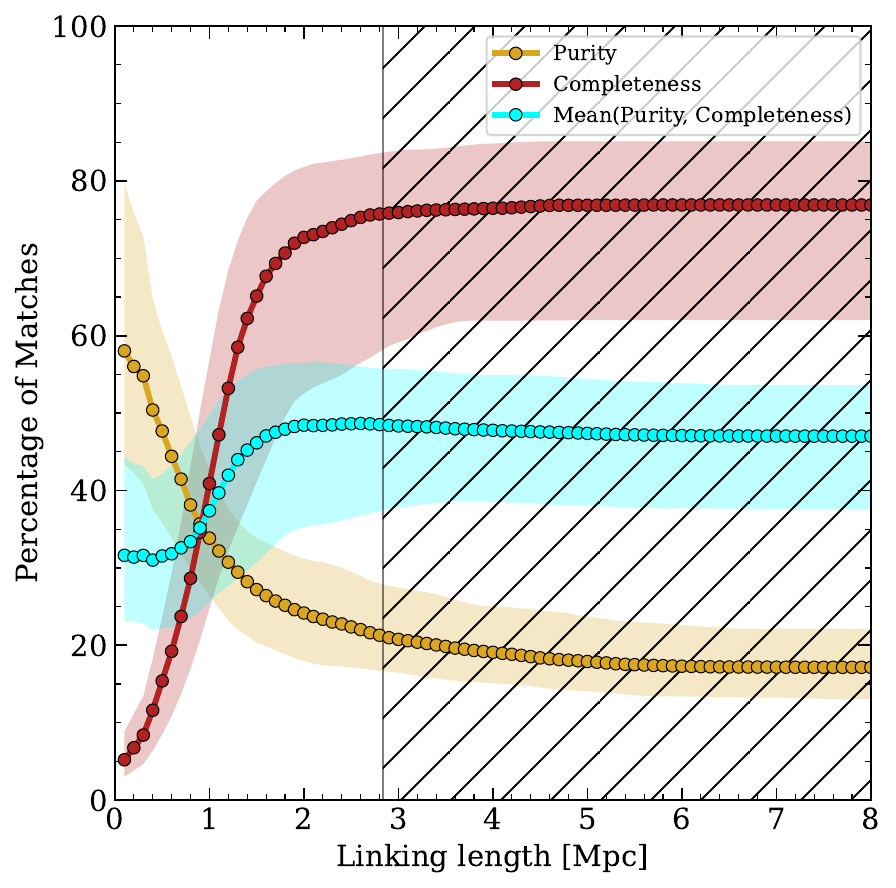}
    \includegraphics[width=0.46\textwidth]{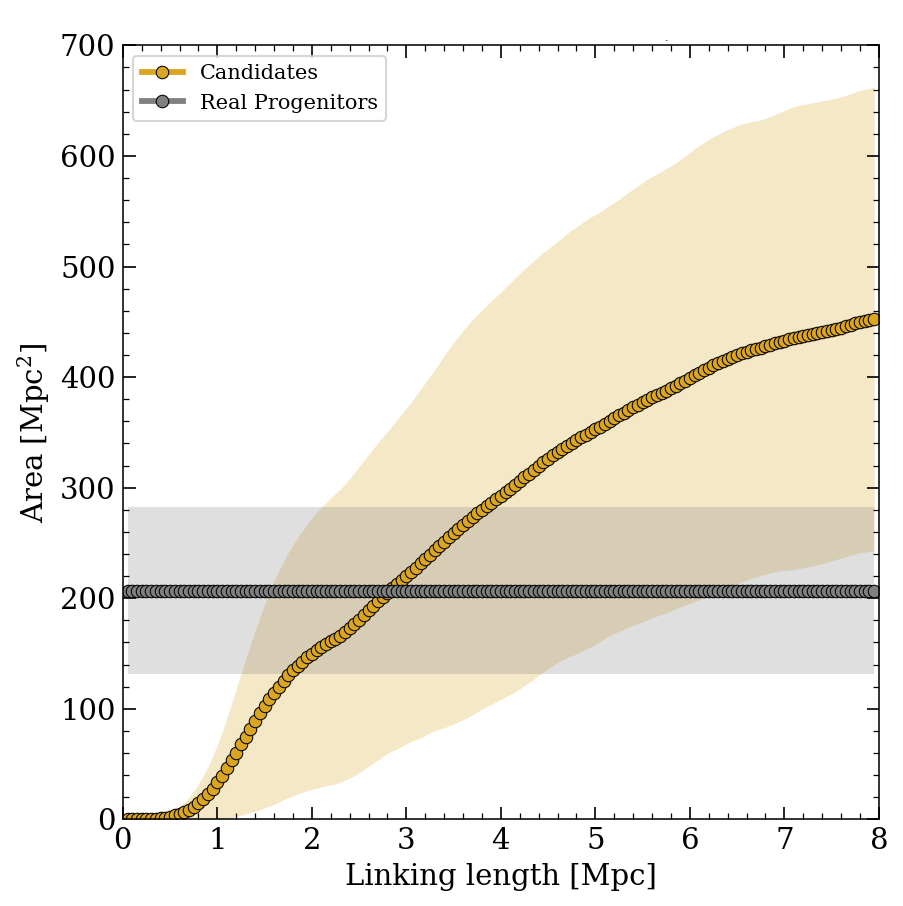}
    \caption{
        One-dimensional Step~1 diagnostic for the representative case at $z=2$ using galaxies with $M_\star>10^{10}\modot$, fixing $n_{\rm n}=7$ and $\alpha=1.0$, and varying the linking length $\ell_1$. The left panel shows purity, completeness, and meanPC averaged over the analysed cluster regions and projected views; shaded bands indicate the corresponding scatter across the sample. The right panel compares the projected area of the \prodigy\ candidate region with the projected area of the extended truth population. The vertical line marks the transition where the mean area ratio satisfies $\langle A_{\rm ratio}\rangle=1$, and the hatched region indicates forbidden configurations with $\langle A_{\rm ratio}\rangle>1$. This figure illustrates why area information is required in addition to membership-based metrics: increasing $\ell_1$ generally increases completeness, but it also expands the candidate region and can lead to inefficient, overly large detections.
    }
    \label{Fig:Appendix:Step1Slices}
\end{figure*}

Figure~\ref{Fig:Appendix:Step1Slices} provides a complementary one-dimensional view of the Step~1 optimisation discussed in Sect.~\ref{Sec:Results:Step1}. In this diagnostic, we fix the nearest-neighbour parameter and overdensity factor to $n_{\rm n}=7$ and $\alpha=1.0$, respectively, and vary only the linking length $\ell_1$. We show the same representative case used in Sect.~\ref{Sec:Results:Step1}, namely the 7K-GIZMO catalogues at $z=2$ with $M_\star>10^{10}\modot$.

This one-dimensional slice makes explicit the mechanism behind the allowed and forbidden regions shown in Fig.~\ref{Fig:Results:Step1Grid}. At small $\ell_1$, the candidate region is too compact and the completeness is low. As $\ell_1$ increases, the algorithm connects more of the extended protocluster structure and the completeness rises. However, the candidate area also grows, and beyond a certain linking length it becomes larger than the projected area of the truth population. These configurations are classified as forbidden. The optimal solutions therefore occur near the transition where meanPC is high while the candidate area remains comparable to, or smaller than, the truth area.

\section{Projection dependence of Step~1 diagnostics}
\label{App:ProjectionDependence}

Figure~\ref{Fig:Appendix:Step1ProjectionDependence} compares the one-dimensional Step~1 performance obtained when using the three orthogonal projected views. The diagnostic is constructed for the same representative case used in Sect.~\ref{Sec:Results:Step1}, fixing $n_{\rm n}=7$ and $\alpha=1.0$ and varying only the linking length $\ell_1$. The curves show that the behaviour of purity, completeness, and meanPC is very similar across projections, and that the transition between allowed and forbidden configurations occurs at comparable values of $\ell_1$. This indicates that projection effects do not introduce a systematic bias in the mean performance of \prodigy\ for this representative case, although they contribute to the scatter across individual regions.

\begin{figure}[htbp]
    \centering
    \includegraphics[width=0.46\textwidth]{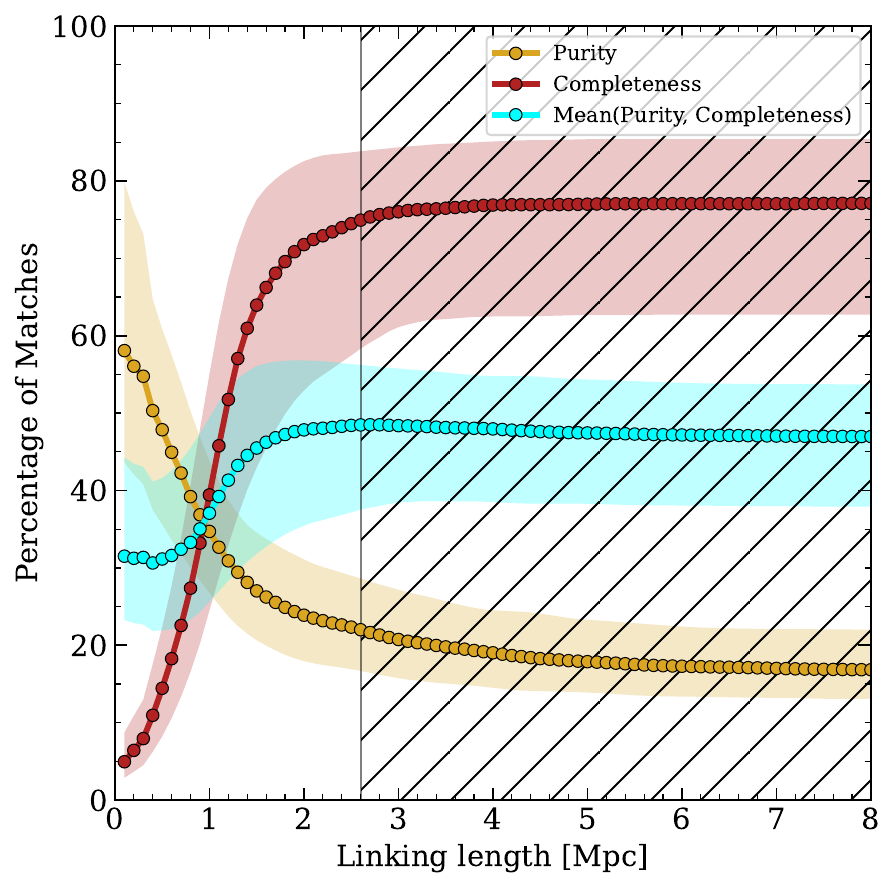}
    \caption{
        One-dimensional Step~1 diagnostic comparing the three orthogonal projected views for the representative case at $z=2$ using galaxies with $M_\star>10^{10}\modot$, fixing $n_{\rm n}=7$ and $\alpha=1.0$. The curves show purity, completeness, and meanPC as a function of the linking length $\ell_1$, averaged over the analysed cluster regions and projected views. Shaded bands indicate the corresponding scatter. The similarity of the curves and of the allowed--forbidden transition shows that the statistical performance of \prodigy\ is only weakly dependent on the projected view.
    }
    \label{Fig:Appendix:Step1ProjectionDependence}
\end{figure}

\section{Step~1 parameter landscape}
\label{App:Step1Landscape}
\begin{figure*}[tbp]
    \centering
    \includegraphics[width=\textwidth]{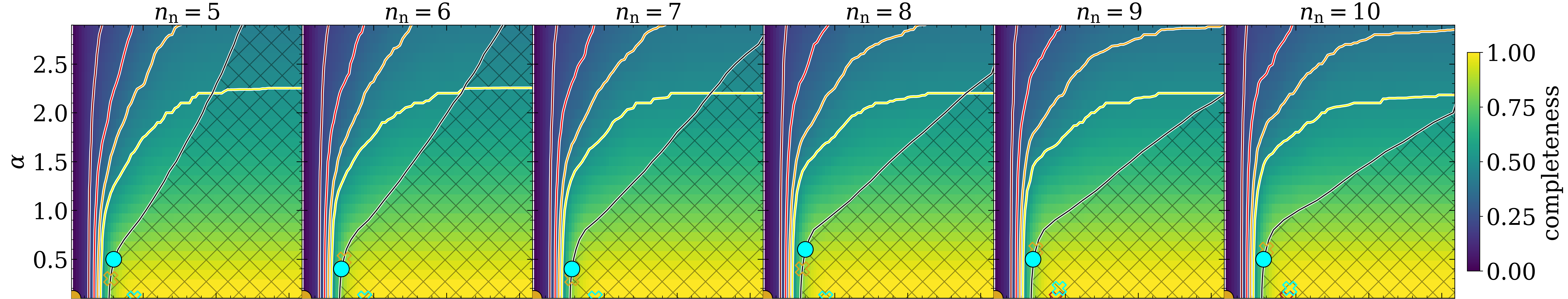} \\
    \includegraphics[width=\textwidth]{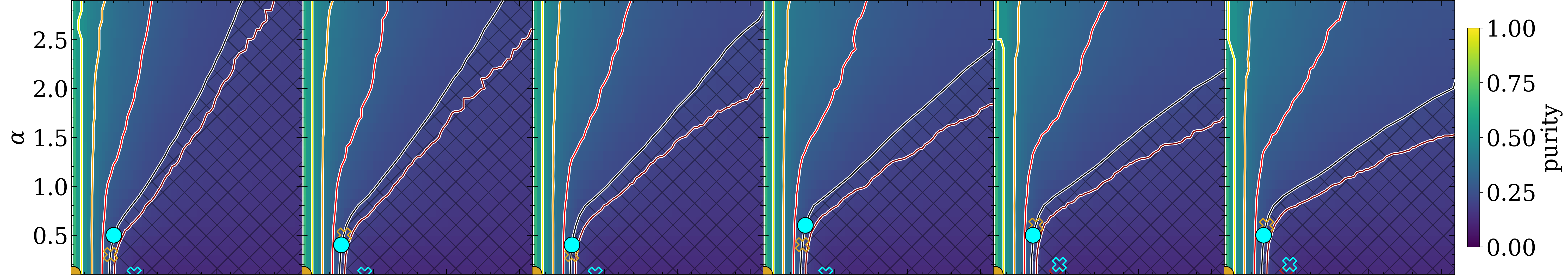}       \\
    \includegraphics[width=\textwidth]{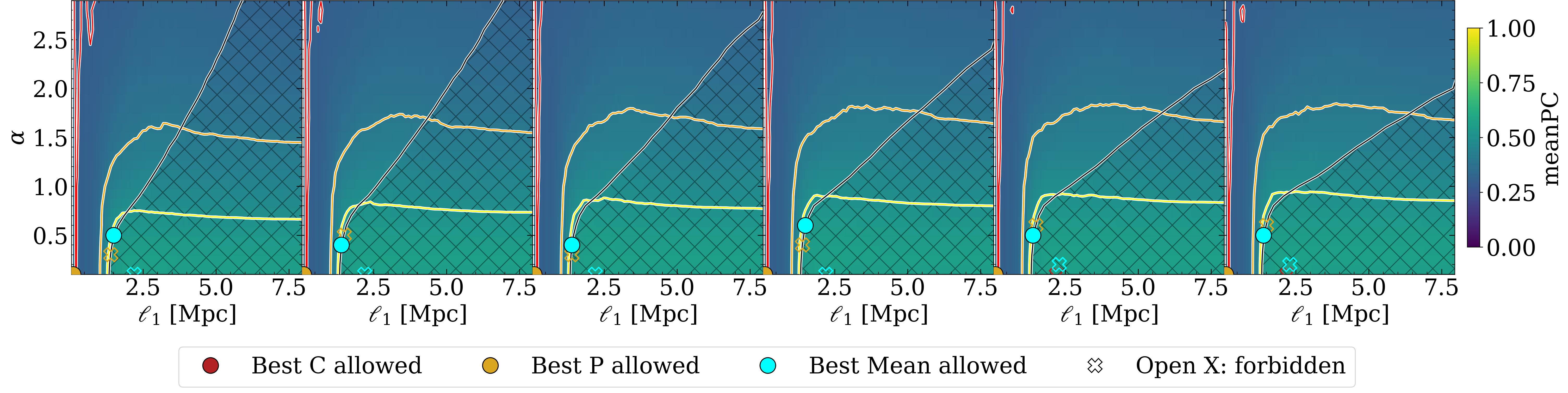}
    \caption{
        Step~1 performance in the \prodigy\ parameter space for the representative case at $z=2$ using galaxies with $M_\star>10^{10}\modot$. Columns correspond to different values of the nearest-neighbour parameter $n_{\rm n}$, while the horizontal and vertical axes show the Step~1 linking length $\ell_1$ and the overdensity factor $\alpha$, respectively. The three rows show completeness, purity, and meanPC, averaged over the analysed cluster regions and projected views. Coloured contour lines indicate metric levels of $0.5$, $0.4$, $0.3$, and $0.2$, shown in yellow, orange, red, and brown, respectively. Hatched regions indicate forbidden configurations, defined by $\langle A_{\rm ratio}\rangle>1$, where the candidate area exceeds the projected area of the extended truth population. The markers indicate the best allowed configurations according to completeness, purity, and meanPC, while open markers indicate optima that would lie in the forbidden region. The figure illustrates the main Step~1 trade-off: permissive configurations recover a larger fraction of the extended protocluster but become less pure and more spatially extended, whereas restrictive configurations are cleaner but less complete.
    }
    \label{Fig:Results:Step1Grid}
\end{figure*}

We now quantify the performance of Step~1 in recovering extended protocluster regions. In this subsection, we focus on a representative case in order to illustrate how the Step~1 optimisation behaves in the three-dimensional parameter space of \prodigy. The systematic dependence on redshift, projection, and stellar-mass threshold is reported in Sect.~\ref{Sec:Results:Dependencies}. Unless otherwise stated, the results shown here correspond to the 7K-GIZMO catalogues at $z=2$, using galaxies with $M_\star>10^{10}\modot$.

Figure~\ref{Fig:Results:Step1Grid} shows the Step~1 performance as a function of the three parameters that control the extended-region detection: the nearest-neighbour scale $n_{\rm n}$, the overdensity threshold factor $\alpha$, and the linking length $\ell_1$. For each value of $n_{\rm n}$, we show the behaviour of completeness, purity, and meanPC in the $(\ell_1,\alpha)$ plane. The hatched region marks configurations that are classified as forbidden according to the area-ratio criterion, namely those for which the mean candidate area is larger than the projected area of the corresponding extended truth population. This representation allows us to identify both the optimal configurations and the broader regions of parameter space where the algorithm behaves robustly.

The expected trade-off between completeness and purity is clearly visible. Low values of $\alpha$ and large values of $\ell_1$ tend to produce highly connected candidate regions that recover a large fraction of the extended truth population. These configurations therefore reach high completeness, but they also include a larger number of interlopers and frequently enter the forbidden region of parameter space. Conversely, high values of $\alpha$ and small values of $\ell_1$ define more restrictive overdense structures. These candidates are generally cleaner and occupy smaller projected areas, but they recover only a limited fraction of the true extended protocluster population. The highest meanPC values are therefore found at intermediate parameter values, where the balance between recovery and contamination is maximised.

The area-ratio criterion is essential for interpreting this optimisation. Without it, configurations with large $\ell_1$ could appear favourable simply because they recover many true members. However, such configurations often do so by selecting projected regions that are larger than the physical footprint of the true protocluster. This behaviour would be inefficient in an observational context, where an excessively large candidate area would imply a larger number of contaminants and a higher cost for spectroscopic follow-up. The allowed region therefore selects configurations that recover the extended protocluster efficiently, rather than configurations that maximise completeness by overextending the candidate region.

The optimisation landscape is also relatively stable across the explored values of $n_{\rm n}$. Although the exact position of the optimal point changes with the nearest-neighbour scale, the global behaviour remains similar: completeness increases towards more permissive thresholds and larger linking lengths, while purity favours more compact and restrictive selections. The best meanPC configurations are not isolated points, but belong to a family of solutions located near the transition between overly compact and overly extended detections. This indicates that Step~1 does not rely on a single fine-tuned parameter choice, but instead admits a robust region of parameter space where the extended protocluster population is recovered with a favourable balance between purity, completeness, and projected area.

A complementary one-dimensional view of the same optimisation problem is presented in Appendix~\ref{App:Step1Slices}. There, we show how purity, completeness, meanPC, and the candidate area vary as a function of $\ell_1$ for fixed $n_{\rm n}$ and $\alpha$. This diagnostic illustrates directly why the area criterion is required: increasing $\ell_1$ generally improves completeness, but also increases the candidate area and eventually moves the solution into the forbidden region.

Overall, Step~1 performs as intended. It is able to recover a large fraction of the extended protocluster population, but only within a restricted part of parameter space where the increase in completeness is not achieved at the expense of excessive contamination or overly large projected areas. The configurations identified here provide the basis for the redshift-, projection-, and mass-threshold-dependent analysis presented in the following subsections.

\section{Step~2 percentile diagnostic}
\label{App:Step2Diagnostic}
\begin{figure}[htbp]
    \centering
    \includegraphics[width=0.48\textwidth]{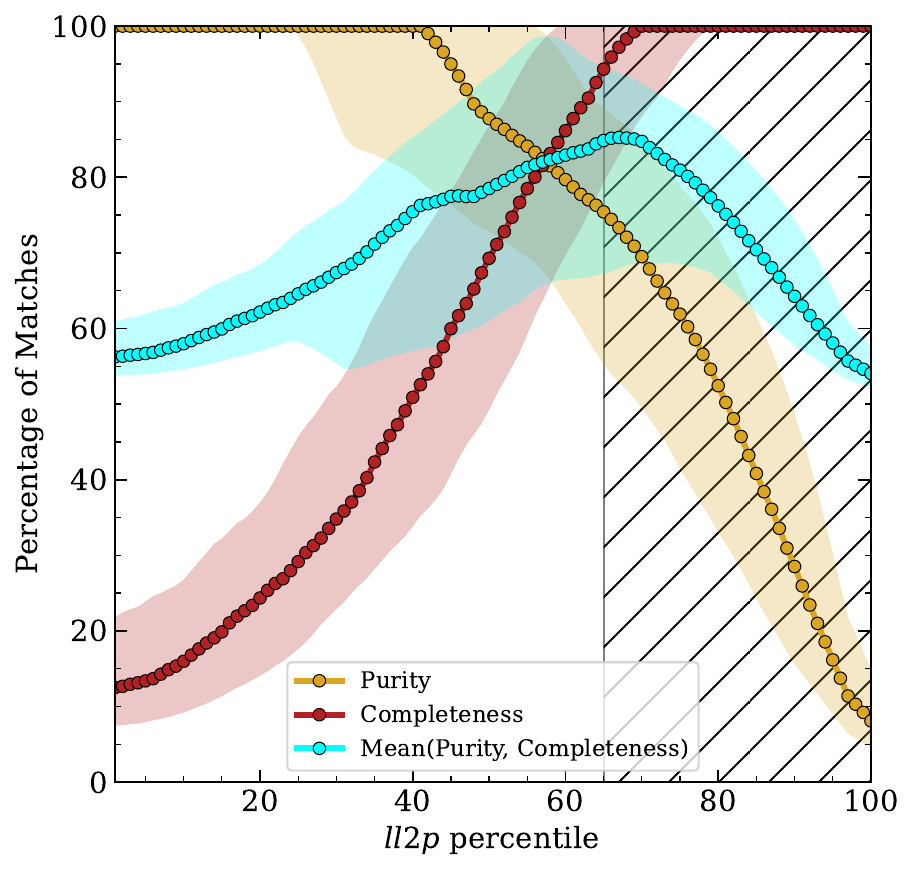}
    \caption{
        Step~2 performance as a function of the percentile parameter $p_2$ for the representative case at $z=2$ using galaxies with $M_\star>10^{10}\modot$. The Step~1 parameters are fixed to the best allowed configuration selected from the extended-region optimisation, and only $p_2$ is varied. The curves show purity, completeness, and meanPC averaged over the analysed cluster regions and projected views; shaded bands indicate the corresponding scatter across the sample. The vertical line marks the transition where the mean area ratio satisfies $\langle A_{\rm ratio}\rangle=1$, and the hatched region indicates forbidden configurations with $\langle A_{\rm ratio}\rangle>1$. Low values of $p_2$ produce compact and highly pure but incomplete cores, while high values of $p_2$ recover more core members at the cost of lower purity and larger projected area.
    }
    \label{Fig:Results:Step2ll2p}
\end{figure}

We now examine the performance of Step~2, which is designed to identify compact and robust protocluster cores within the extended environments recovered by Step~1. As in Sect.~\ref{Sec:Results:Step1}, we focus on the representative case at $z=2$ using galaxies with $M_\star>10^{10}\modot$. The Step~1 parameters are fixed to the best allowed configuration selected from the extended-region optimisation, and we then vary only the Step~2 percentile parameter $p_2$, which controls the adaptive linking length $\ell_2$ defined in Eq.~\ref{Eq:Methods:Step2:ll2}.

Figure~\ref{Fig:Results:Step2ll2p} shows the Step~2 purity, completeness, and meanPC as a function of $p_2$. The trends are physically intuitive. At low values of $p_2$, the adaptive linking length is small, so the algorithm selects only the most compact part of the overdense region. These configurations have very high purity, because most selected galaxies belong to the core truth population, but their completeness is low because a substantial fraction of real core members remains unselected. As $p_2$ increases, the adaptive linking length grows and Step~2 connects a larger fraction of the core. The completeness therefore increases steadily, while the purity decreases as the candidate begins to include additional galaxies outside the core truth set.

The resulting meanPC curve peaks at intermediate values of $p_2$, where the balance between purity and completeness is optimal. This behaviour is analogous to the Step~1 optimisation, but with a different physical target. Step~1 searches for the extended protocluster environment and is therefore allowed to recover a broad, irregular region, whereas Step~2 is evaluated against $\mathcal{T}^{\rm core}_{\mathcal{R}}(z)$ and is expected to remain compact. The hatched region in Fig.~\ref{Fig:Results:Step2ll2p} marks configurations for which the mean candidate area exceeds the projected area of the core truth population. These configurations can reach high completeness, but they do so by selecting regions that are too extended relative to the physical core.

The area constraint is particularly important for Step~2. Since the purpose of this stage is to define a high-confidence core, an overly large candidate would defeat the observational motivation of the method. The best Step~2 solutions are therefore those that maximize meanPC while remaining in the allowed region of parameter space. In the representative case shown here, this occurs near the transition between the allowed and forbidden regimes, where Step~2 preserves a high fraction of the true core population while keeping the candidate area comparable to, or smaller than, the projected core footprint.

Overall, Step~2 performs as intended: it converts the extended Step~1 detection into a more compact candidate associated with the densest part of the protocluster. Relative to the extended-region search, the core-identification step reduces the projected extent of the detection and increases the expected reliability of the selected region, while maintaining a high completeness with respect to the core truth population. This makes Step~2 especially relevant for observational applications, where a compact high-purity region can be used to prioritize spectroscopic follow-up and membership confirmation.

\section{Catalogue construction and projection details}
\label{App:CatalogueDetails}
The following details are retained from the input manuscript to document the catalogue construction underlying Sect.~\ref{Sec:Dataset:GalaxySelection}.

In this work, we use the 7K-GIZMO hydrodynamical galaxy catalogues of {\sc The Three Hundred} Project. These catalogues contain full three-dimensional positions, velocities, host-halo information, and merger-tree links. However, the \prodigy\ detection itself is performed on projected two-dimensional galaxy distributions; the three-dimensional and merger-tree information is used only to define the descendant-based reference populations and to identify the main progenitor halo. The use of the hydrodynamical sample is important because the galaxies are formed self-consistently within the simulation and can therefore be used directly as tracers of the high-redshift protocluster population. All haloes and subhaloes in the 7K-GIZMO runs were identified with the Amiga Halo Finder\footnote{\url{http://popia.ft.uam.es/AHF}} \citep[AHF;][]{knollmann&knebe2009}, which locates density peaks using an adaptive mesh hierarchy and removes gravitationally unbound particles from each candidate structure. For each halo, AHF provides standard spherical-overdensity quantities such as $R_{200{\rm c}}$ and $M_{200{\rm c}}$, defined respectively as the radius enclosing a mean density equal to 200 times the critical density of the Universe and the total mass enclosed within that radius. In the hydrodynamical catalogues, the properties of the stellar, gas, and dark matter components associated with each bound structure are also recorded.

The temporal connection between haloes and subhaloes across the 129 simulation snapshots, spanning redshifts from $z\sim17$ to $z=0$, is built using the {\sc MergerTree} algorithm distributed with AHF. This merger-tree information allows high-redshift galaxies and their host subhaloes to be connected to their $z=0$ descendants, and is used in this work to construct the descendant-defined reference populations introduced in Sect.~\ref{Sec:Dataset:GroundTruth}.

For each simulated galaxy, the catalogue includes intrinsic physical properties such as stellar mass, gas mass, star formation rate, metallicity, position, velocity, and host-halo information.

In this paper, however, the protocluster detector is calibrated using stellar-mass-selected galaxy samples rather than colour- or flux-selected catalogues. Unless otherwise stated, our fiducial tracer population is defined by selecting galaxies with stellar masses above $M_\star > 10^{10}\modot$, while additional samples with thresholds of $M_\star > 10^{9}\modot$ and $M_\star > 10^{8}\modot$ are also considered. These thresholds allow us to assess how tracer depth affects the performance of \prodigy. In particular, the $M_\star>10^8\modot$ sample provides a more complete sampling of the simulated progenitor population, allowing us to explore the algorithm's behaviour with less restrictive stellar-mass selections. However, this threshold is not intended to represent a stellar-mass-complete observational sample at $z=3$--4, where survey selection effects must be considered. The dependence of the \prodigy\ performance on the adopted stellar-mass selection is explored explicitly in Sect.~\ref{Sec:Results}.

The use of three orthogonal projections allows us to quantify how the performance of \prodigy\ depends on the viewing direction. This is important because protoclusters are intrinsically extended and anisotropic, with observations showing that they can be embedded in coherent filamentary large-scale structures \citep{matsuda2005,muldrew2015,overzier2016}. Their projected appearance may therefore change substantially depending on the line of sight. Projection can either enhance the apparent significance of an overdensity by superposing physically separated structures, or dilute a genuine protocluster if its member galaxies are spread over a large projected area. By applying the same detection procedure to the $xy$, $xz$, and $yz$ views, we obtain a controlled estimate of the projection-induced variance in purity, completeness, meanPC, and area ratio.

\clearpage
\onecolumn
\section{Observational footprints and reference mapping}
\label{App:ObservationalFootprints}
The observational overlay in Fig.~\ref{Fig:Results:BestParameters:Reff} uses the supplied 172-entry compilation. Table~\ref{Tab:ObservationalFootprints} maps every legend group to its literature sources; repeated systems can represent different spatial selections. Ref identifiers refer to legend groups, not to distinct or independent objects. The accompanying machine-readable compilation retains all supplied values and proxy notes. Most of the high-redshift aperture selections and the six volume-derived proxies were collected through the compilation of \citet{baxter2025}; the latter originate in \citet{shah2024}. We distinguish this secondary compilation from the original observational studies.

For simulations, $R_{\rm eff}=\sqrt{A/\pi}$ is defined from the projected population footprint. For observations, a reported aperture radius is retained as an aperture proxy, a field area gives $\sqrt{A_{\rm field}/\pi}$, and an extent or volume-derived radius retains its supplied geometric interpretation. Physical and comoving units are related by $R_{\rm com}=(1+z)R_{\rm phys}$; angular conversion involves the angular-diameter distance of the adopted cosmology. The supplied observational conversions are retained without imposing a uniform cosmology or adding unreported uncertainties. Lensed source-plane sizes are not interchangeable with image-plane observing coverage. The CHANCES and GOGREEN points use virial-radius measurements, rather than full survey or progenitor-environment footprints.

\begingroup
\footnotesize
\setlength{\tabcolsep}{4pt}
\begin{longtable}{lp{3.2cm}p{5.0cm}p{7.0cm}}
    \caption{Legend groups, footprint conventions, and observational sources for Fig.~\ref{Fig:Results:BestParameters:Reff}.}\label{Tab:ObservationalFootprints}\\
    \hline
    ID & Legend group & Supplied proxy & Sources \\
    \hline
    \endfirsthead
    \multicolumn{4}{l}{Table~\ref{Tab:ObservationalFootprints}, continued.}\\
    \hline
    ID & Legend group & Supplied proxy & Sources \\
    \hline
    \endhead
    \hline
    \endfoot
    Ref1 & Planck Herschel & Aperture & \citet{polletta2021} \\
    Ref2 & PKS1138 & Aperture; source link provisional & \citet{pentericci1998} \\
    Ref3 & BOSS & Different selected apertures & \citet{shi2021,cai2017} \\
    Ref4 & HS1700 & Aperture & \citet{lacaille2019} \\
    Ref5 & PCL1002 & Aperture & \citet{casey2015} \\
    Ref6 & 4C23.56 & Compact aperture & \citet{lee2017} \\
    Ref7 & USS1558 & Aperture & \citet{perezmartinez2024} \\
    Ref8 & Baxter compilation & Volume-derived footprint proxies & \citet{baxter2025,shah2024} \\
    Ref9 & HS1549 & Aperture & \citet{lacaille2019} \\
    Ref10 & MRC & Radio-galaxy apertures & \citet{venemans2007} \\
    Ref11 & VUDS & Selected projected regions & \citet{cucciati2014,lemaux2014} \\
    Ref12 & TNJ & Radio-galaxy apertures & \citet{venemans2007,overzier2008} \\
    Ref13 & SSA22 & Selected footprint & \citet{steidel1998} \\
    Ref14 & MAGAZ3NE & Selected footprint & \citet{mcconachie2022} \\
    Ref15 & HSC SSP & Apertures; legend grouping & \citet{toshikawa2016,toshikawa2025} \\
    Ref16 & PC217 & Selected core aperture & \citet{dey2016} \\
    Ref17 & DRC & Compact proxy; value provisional & \citet{oteo2018} \\
    Ref18 & SPT & Compact selected footprint & \citet{miller2018} \\
    Ref19 & HDF & Selected footprint & \citet{calvi2023} \\
    Ref20 & JADES & Selected region; 42/21 arcsec apertures & \citet{helton2024,bogdan2026} \\
    Ref21 & Subaru & Selected footprints & \citet{harikane2019} \\
    Ref22 & JWST & ODz5p8 aperture & \citet{morishita2025} \\
    Ref23 & COSMOS & Selected footprint & \citet{brinch2024} \\
    Ref24 & SDF & Selected footprint & \citet{toshikawa2014} \\
    Ref25 & HSC & Selected footprint & \citet{higuchi2019} \\
    Ref26 & LAGER & Large selected aperture & \citet{hu2021} \\
    Ref27 & ASPIRE & Mosaic area; ambiguous extra label & \citet{champagne2025} \\
    Ref28 & SMACS0723 & Lensed source-plane radius & \citet{laporte2022} \\
    Ref29 & A2744 & Lensed source-plane radius & \citet{morishita2023_glass} \\
    Ref30 & ClG J0218 & Environment and central aperture & \citet{hatch2016,santos2014} \\
    Ref31 & 7C1756 & Membership aperture & \citet{casasola2018} \\
    Ref32 & XMMXCS J2215 & Field-area equivalent and $R_{200}$ & \citet{hayashi2011} \\
    Ref33 & XMM-LSS & Membership aperture & \citet{papovich2010} \\
    Ref34 & ClJ1449 & Central and environmental apertures & \citet{coogan2018,smith2019} \\
    Ref35 & VUDS Hyperion & Total field-area equivalent & \citet{cucciati2018_hyperion} \\
    Ref36 & GOGREEN & $R_{200\mathrm{c}}$ cluster anchors & \citet{old2020_gogreen} \\
    Ref37 & CHANCES Low-z & $R_{200}$; not $5R_{200}$ coverage & \citet{sifon2025_chances} \\
    Ref38 & CHANCES Evolution & $R_{200}$; not $5R_{200}$ coverage & \citet{sifon2025_chances} \\
    Ref39 & Hyperion components & Seven projected component proxies & \citet{cucciati2018_hyperion} \\
\end{longtable}
\endgroup

\noindent\textit{Source corrections and qualifications.} Publication metadata resolves the input labels ``Bogdan2025'' to \citet{bogdan2026}, the ODz5p8 ``Morishita2024'' entry to \citet{morishita2025}, and the A2744 $z=7.88$ entry to \citet{morishita2023_glass}. The XMM-LSS membership aperture is described by \citet{papovich2010}, rather than a first-author Brodwin paper. Cl~J1449 uses the complementary central and environmental studies of \citet{coogan2018,smith2019}. The extra ASPIRE entry labelled ``ClJ1001'' in the input is ambiguous and is not used to interpret a separate system; the explicitly identified ASPIRE mosaic is the preferred comparison. The PKS1138 aperture-source association and the unusually small supplied DRC proxy remain provisional, so neither supports a quantitative object-specific size claim. Hyperion component radii are supplied projected proxies derived from the published component dimensions; their numerical convention differs from the simulation area-equivalent radius. These qualifications concern interpretation of the supplied overlay, not a reconstruction of the observational or simulation analysis.
\twocolumn
\end{appendix}

\end{document}